\documentclass[12pt,preprint]{aastex}

\def\F0{F_{\rm 0}}
\def\t0{t_{\rm 0}}
\def\NE{N_{\rm E}}
\def\FE{F_{\rm E}}
\newcommand{\ltsima} {$\; \buildrel < \over \sim \;$}
\newcommand{\gtsima} {$\; \buildrel > \over \sim \;$}
\newcommand{\lta} {\lower.5ex\hbox{\ltsima}}
\newcommand{\gta} {\lower.5ex\hbox{\gtsima}}

\slugcomment{Date: \today}

\begin{document}

\title{The Cooling Behavior of Thermal Pulses in Gamma-Ray Bursts}

\author{Felix Ryde}

\affil{Stockholm Observatory, AlbaNova, SE-106 91 Stockholm,
Sweden}

\begin{abstract}

We discuss gamma-ray bursts (GRBs) that have very hard spectra,
consistent with black-body radiation. Several emission components
are expected, on the basis of theoretical considerations, to be
visible in the gamma-ray band, mainly non-thermal emission from
cooling, relativistic electrons and thermal emission from a wind
photosphere. We find that the pulses we study are consistent with
a thermal, black-body radiation throughout their duration and that
the temperature, $kT$, can be well described by a broken power-law
as a function of time, with an initially constant or weak decay
($\sim 100$ keV). After the break, most cases are consistent with
a decay with index $-2/3$. A few of the pulses have a weak
non-thermal component overlayed the thermal one, and are better
fitted with a combination of a thermal and a non-thermal
component. We further demonstrate that such a two-component model
can explain the whole time-evolution of other bursts, that are
found to be only initially thermal and later become non-thermal.
The relative strengths between the two components vary with time
and this is suggested to, among other things, account for the
change in the modelled low-energy power-law slope that is often
observed in GRBs. The secondary, non-thermal components are
consistent with optically-thin synchrotron emission in the cooling
regime. We interpret the observations within a model of an
optically thick shell (fireball) that expands adiabatically. The
slow, or constant, temperature decrease is from the acceleration
phase, during which the bulk Lorentz factor increases, and the
faster temperature decay is reached as the flow saturates and
starts to coast with a constant speed. We also discuss a
Poynting-flux model, in which the saturation radius is reached
close to the photosphere. Even though these observations cannot
tell these models apart, the latter has several attractive
advantages. The GLAST satellite will be able to clarify and
further test the physical setting of similar thermal pulses.

\end{abstract}


\section{Introduction}

The radiation from gamma-ray bursts (GRBs) is assumed to arise in
a highly relativistic outflow from a collapsing massive progenitor
star. This outflow could either be in a form of a strongly
magnetized wind, where most of the energy is transported in
magnetic fields, or in a low-baryon-load 'fireball' where the
energy is transported mainly as kinetic energy. In both cases the
outflow is initially optically thick and will emit thermal
radiation. As the flow expands the typical photon energy and the
number density will decrease. At a certain distance from the
progenitor it will become optically-thin to pair production and
Compton scattering off the free electrons associated with the
baryons in the outflow, and the radiation decouples from the
matter and escapes to an observer at infinity. The plasma then
emits a strong flash of thermal black-body radiation. Beneath this
photosphere the radiation created by the dissipation of energy is
transformed back into the kinetic energy of the flow. As the
outflow becomes transparent, non-thermal radiation processes such
as synchrotron and/or inverse Compton processes can radiate away
the dissipated energy, by transforming it into the gamma-rays. In
the {\it fireball model}, the wind has a variable initial
distribution of Lorentz factors, often thought of as individual
shells moving with various speeds. In this outflow shocks are
created, at a typical distance of $\sim 10^{14}$ cm from the
center of the progenitor, which dissipate the energy and
accelerate the electrons which cool and radiate. The radiation
could also stem from shocks created as the wind interacts with the
external medium surrounding the progenitor. On the other hand, in
the magnetic-energy dominated, or {\it Poynting flux models}, the
magnetic field energy is assumed to be dissipated locally through
reconnections, and is then transformed into $\gamma$-rays by the
accelerated electrons.

The observed spectrum from a GRB is therefore expected to be a
superposition of the thermal black-body and the
synchrotron/inverse Compton emissions generated in the
optically-thin environments. Depending on the initial conditions
of the outflow, the radiative efficiencies in the shocks and/or
reconnection rates, these components will be of different
importance \citep{Mesz,DS}. For instance, \citet{DM} calculated
the expected photospheric black-body luminosity and temperature
for various fireball models. For the standard model  they found
that the internal energy will still be large at typical
photosphere radii, and thus the thermal component of the spectrum
will be very hot and luminous. Depending on the observed energy
band it could even be dominating the spectrum, even when the
radiative efficiency in the shocks is very high. Similarly,
\citet{DS} found in numerically modelling of strongly magnetized
winds from GRBs that, for certain initial conditions, the amount
of energy emitted as thermal radiation can indeed be large.

Most observed GRB light curves are dominated by a non-thermal
radiation \citep{preece00} and a large fraction of these are
consistent with various versions of an optically-thin, intrinsic
synchrotron spectrum, that is, consistent with a photon index of
$\alpha < -2/3$ (see eq. [\ref{eq:band}] below for the definition
of $\alpha$). This radiation naturally stems from energy
dissipation in optically-thin regions and is usually interpreted
as the signature of, for instance, the internal shocks.
Furthermore, the power-law below the peak energy, i.e. $\alpha$,
is variable for more than half of all bursts, and of these, it
softens in general \citep{crider}. However, a notable fraction of
the time-resolved spectra are harder than optically-thin
synchrotron spectra, that is, inconsistent with $\alpha < -2/3$,
the so-called line-of-death for the synchrotron model \citep{lod}.
It is mainly spectra early on in the pulse light-curve that are
the hardest. Aside from theoretical motivations mentioned above
[see, e.g., \citet{Mesz, DM, DS, blinnikov}], several authors have
already proposed a thermal origin as an interpretation of the very
hardest {\it observed} time-resolved spectra at early times, for
instance \citet{crider, ghirlanda}. However, no conclusive picture
has yet emerged explaining these 'hard' spectra and especially
their spectral evolution. In this paper, we identify a few bursts
which are consistent with black body emission throughout the
burst, which has not been reported before. These are presented in
\S \ref{sec:sample} and their behavior is analyzed in \S
\ref{sec:analysis}. We interpret these as bursts in which the
initial conditions are such that the thermal photospheric emission
component dominates over the optically-thin emission from the
dissipated free energy. We can thus study the temperature behavior
by modelling the spectrum with an intrinsic black-body spectrum.
We also analyze a few pulses in which a power-law component in the
spectrum becomes more and more dominant, thus accounting for the
initially hard bursts and their spectral evolution. We discuss
this interpretation in \S \ref{sec:interpretation} and we conclude
our discussion in  \S \ref{sec:disc}.

\section{The Sample of Thermal Pulses and the Analysis Method}
\label{sec:sample}

To find bursts that are dominated by a thermal emission, we search
among bursts that have a prominent, single pulse(s). The
motivations for this is manyfold. First, the hardest spectra that
have been reported are time-resolved spectra at the beginning of
simple pulse structures (see, e.g., \citet{ghirlanda}). Second, in
the numerical study of single-pulsed GRBs within the standard
internal shock model, \citet{DM} concluded that the thermal
component should be in most cases at least as bright as the
non-thermal component for such pulses.  Third, in several other
theoretical/numerical modelings (e.g., \citet{remo01} and
\citet{Lyu}) a single, thermal, precursor-pulse ('proper-GRB' in
\citet{remo01}) is found to preceed the more variable main burst
emission. Fourth, from a theoretical point of view the variability
of a burst is closely related to the location of the photosphere
(e.g., \citet{KRM}). The further out the photosphere is, for
instance due to low bulk Lorentz factors, the less efficient are
the internal shocks and the less variable are the light curves. In
these cases the thermal photosphere could be expected to be
relatively stronger. Fifth, in the analysis in McQuinn, Ryde, \&
Petrosian (2004, in prep.) we conclude that the curvature effect
is more prominent for hard spectral bursts, such we here interpret
as thermal, compared to non-thermal bursts. This effect is due to
differences in light-travel times, and Lorentz boosts for photon
emitted from a curved surface. It has a strong effect on the light
curves, making them smoother. All these arguments point to that
pure thermal emission is expected to be a smooth pulse, or at
least dominated by one, while the non-thermal emission from the
internal shocks and/or reconnections is expected to be highly
variable with complicated light curves. This therefore motivated
us to search for photospheric bursts in the catalogue of GRB
pulses by F. Ryde et al. (in prep.). These bursts are a subsample
of the {\it CGRO} BATSE catalogue and comprise strong, single
pulses. The selection criterion that was used for these pulses
required the peak flux to be greater than 1.0 photons cm$^{-2}$
s$^{-1}$ on a 256 ms timescale. Of these, bursts were selected
that exhibited clean, single-peaked events, or in the case of
multi-peaked bursts, pulses that were well separated and
distinguishable from each other. The catalogue is limited to
pulses with durations longer than 2 s (full width--half max). In
the search we did not assume any functional description of the
pulse profile in order to eliminate any preconceived idea of what
the fundamental pulse shape should be. The final sample consisted
of 76 pulses within 68 bursts. This sample has also been used in
other connections, for instance, in \citet{KRL}.

The data were gathered by BATSE's Large Area Detectors (LADs),
which flew on the {\it Compton Gamma-Ray Observatory (CGRO)}. We
used mainly the high-energy resolution burst (HERB) data
\citep{Fish89}. The HERB data have 128 energy channels covering
approximately 10 keV to several MeV, with sub-second
time-resolution. A background estimate was made using the HER
data, which consist of low (16--500~s) time resolution
measurements that are stored between triggers.  The light curve of
the background, during the outburst, was modelled by interpolating
these data, roughly 1000 s before and after the trigger, with a
second or third order polynomial fit. The data, as well as the
background fit coefficients were obtained from the CGRO Science
Support Center (GROSSC) at Goddard Space Flight Center through its
public archives. The central part of the analysis was performed
with the RMFIT package, version 1.0b1 (Malozzi et~al. 2000) and
the spectral fitting was done using the MFIT package, version 4.6,
running under RMFIT.  We always chose the data taken with the
detector which was closest to the line-of-sight to the location of
the GRB, as it has the strongest signal. The broadest energy band
with useful data was selected, which is often approximately
25--1900 keV, as the lowest $\sim$ 8 channels of the HERB data are
usually not useful. Although BATSE's spectroscopic detectors (SDs)
can gather enough data to produce spectra on a 0.128 s timescale,
the data usually have to be integrated to get an acceptable fit to
a given spectra model. This reduces the resolution of the
time-history significantly. The background-subtracted {\it photon}
spectrum, $\NE(E)$, was determined using a forward-folding
technique. A spectral model was folded through the
detector response matrix and was then fitted by minimizing the
$\chi ^2$ (using the Levenberg-Marquardt algorithm) between the
model count spectrum and the observed count spectrum, giving the
best-fit spectral parameters and the normalization.

In our search for photospheric candidates, we fitted the spectra
using a Planck function described by equation (\ref{eq:BBph})
below. The bursts that were found to be reasonably well fitted by
black bodies were selected: GRB930214 (BATSE trigger \#2193),
GRB941023 (\#3256), GRB951228 (\#4157), GRB971127 (\#6504),
GRB990102 (\#7295). These are presented in Table~\ref{tab:one},
where they are denoted by both their GRB names and BATSE trigger
numbers. We also present, in  Figure \ref{fig:lc1}, the 64 ms
resolution count light-curves. These data have the maximal
time-resolution, but have only  four energy channels. The peak
times are also given in the table.

To characterize the {\it time-integrated} spectra of the pulses,
we model them with the empirical function (the 'GRB-function';
\citet{band93}):
\begin{equation}
N_{\rm E}(E) = \left\{ \begin{array}{ll}
            C \ E^{\alpha}
e^{-E/E_{\rm 0}} & \mbox{if $(\alpha-\beta) E_{\rm 0} \geq E$}\\
            C' E^{\beta} & \mbox{if $(\alpha-\beta)
            E_{\rm 0} < E$ ,} \end{array} \right.
\label{eq:band}
\end{equation}
\noindent where $E$ is the energy in keV, $E_{\rm 0}$ is the
$e-$folding energy, $\alpha$ and $\beta$ are the asymptotic power
law indices, $C$ the amplitude, and $C'$ has been chosen to make
the photon spectrum, $N_{\rm E}(E)$, a continuous and a
continuously differentiable function through the condition
\begin{equation}
C'=C \left[ (\alpha-\beta)E_{\rm 0} \right]^{\alpha-\beta}
e^{-(\alpha-\beta)}.
\end{equation}
The peak energy, $E_{\rm pk}$, at which the $EF_{\rm E}$-spectrum
($ \FE \equiv  E \NE $) is at its  maximum, is used as a measure
of the spectral hardness instead of $E_{\rm 0}$. They are related
by $E_{\rm pk}= (2+\alpha) E_{\rm 0}$. The properties of the
time-integrated spectra are given in table \ref{tab:two}. In
particular, we want to draw attention to the low-energy power-law
indices, $\alpha$, which, in general, are hard, that is, larger
than $-2/3$, but they are all lower than, for instance $+1$, the
Rayleigh-Jeans spectrum of a black body (see \S
\ref{sec:analysis}). One case, trigger 3256, even has a slope that
is softer than $-2/3$, even though the time-resolved spectra are
hard. The instantaneous and the time-integrated spectra are, in
general, different, since there is significant spectral evolution.

We will later also discuss other bursts that have previously been
found to have hard spectra, consistent with black bodies, but only
in the initial phase of the burst. These quickly become
non-thermal:
910927 (\#829), 911118 (\#1085), GRB
970111 (\#5773) and GRB980306 (\#6630). These pulses are shown in
Figure \ref{fig:lc2}.

\section{Spectral Modelling}
\label{sec:analysis}

\subsection{Thermal and non-thermal spectra}

The {\it time-resolved, thermal} spectra of the pulses, identified
in the previous section, are modelled by a Planck function, which
is the characteristic spectrum from an optically-thick medium in
which the mean-free-path of a photon is short and the radiation is
completely coupled to the matter. The background-subtracted photon
spectrum, $ \NE(E,t)$, for each time bin of the observations, is
thus modelled by \citep{planck01}
\begin{equation}
\NE (E,t) = A(t) \,\, \frac{ E^2}{exp[E/kT(t)]-1} = A\,(kT)^2\,
\frac{x^2}{(e^x - 1)}, \label{eq:BBph}
\end{equation}
\noindent where $x \equiv E/k T$ and $kT$ is the color temperature
of the black-body in keV, and $k$ is the Boltzmann constant
($8.617 \times 10^{-8}$ keV/K). The fitted parameters are thus the
normalization $A$ and temperature $kT$, and  the quality of the
fit is given by a reduced $\chi^2$-value ($\equiv \chi^2 _{\nu}$).
For small photon energies compared to the temperature ($E<<kT$)
the Planck function approaches the Rayleigh-Jeans Law $\NE (E)
\propto E$, and at large energies ($E>>kT$) it approaches the Wien
law $\NE (E) \propto E^2 \exp (-E/kT)$. In the discussion below we
will denote the spectrum below $kT$ as the Rayleigh-Jeans portion
of the spectrum and similarly, the Wien portion of the spectrum is
that above $kT$. Strictly speaking, however, the asymptotic laws
are not reached until several decades away from $kT$.  This is
illustrated in figure \ref{fig:planck}, in which a Planck curve
(solid line) is shown in a $E F_E$-representation. The energy
interval 20 keV -- 2 MeV is also indicated. It is clear from the
figure that the Rayleigh-Jeans power-law is barely reached within
the first decade below the peak. This aspect is important to bear
in mind when observations in a finite band is interpreted.

{\it Non-thermal}  GRB spectra are often interpreted as being the
result of synchrotron emission from a distribution of relativistic
electrons. In the simplest version of the model the radiation is
emitted from an optically-thin  plasma in which the electron
pitch-angles are isotropically distributed \citep{pac}. In
relativistic shocks, electrons are accelerated, through various
mechanisms, into a power-law distribution in Lorentz factors, say
with index $p$ and within an interval of Lorentz factors
[$\gamma_{\rm min}\leq\gamma\leq\gamma_{\rm max}$];
$f(\gamma)\propto \gamma^{-p}$. If the electrons cool on a very
short time-scale, this distribution will steepen by unity;
$-(p+1)$. For instance, for the Fermi type of particle
acceleration in relativistic shocks $p \sim 2.2$ \citep{mesrev}.
However, the nature of the actual acceleration mechanism that
takes place in GRBs is not well understood.

In all cases, the distribution of electrons will give rise to a
characteristic photon spectrum  with a peak at an energy that is
related to the minimum value of the electron Lorentz factors,
$E_{\rm p} \propto \gamma _{\rm min} ^2 h \nu_B$, where $\nu_{\rm
B} = eB_{\perp}/2\pi m_{\rm e} c$ is the gyrofrequency in a
magnetic field with a mean perpendicular component $B_{\perp}$.
Below this frequency the spectrum approaches a power law with
index $\alpha = -2/3$, while at higher energies the power law will
have an index of $\beta = -(p+1)/2$ (cooling effects are
negligible) or $\beta = -(p+2)/2$ (cooling spectrum). The main
contribution to the low-energy spectrum is from the lowest energy
electrons emitting with their characteristic spectrum. The value
of $\alpha = -2/3$ is therefore the largest (hardest) value that
optically-thin synchrotron-emission can yield. To approximately
model such spectra various versions of broken power-laws have been
used, for instance the GRB-function [eq. (\ref{eq:band})]. In \S
\ref{sec:sp} below we will characterize the non-thermal
contribution to the spectrum by a single power-law, which is the
best allowed by the data, but nevertheless informative.

\subsection{Evolution of the Black-body Spectra}
\label{sec:evol}

\subsubsection{Planck Model Fits}

The results of the spectral modelling of the five triggers in the
sample with a black body model is depicted in Figures
\ref{fig:2193} to \ref{fig:7295}. Here, the time-resolved {\it
photon-flux} spectral data  are shown, as well as the
corresponding best-fit Planck function (eq. [\ref{eq:BBph}]). The
crosses correspond to the deconvolved data points in photons
cm$^{-2}$ s$^{-1}$ keV $^{-1}$ and the spectra are rebinned to
have a SNR=1 to make the plots clearer. Ten characteristic
time-bins were selected for each case to illustrate the time
evolution of the spectra during the pulses.

The first conclusion that can be drawn from the figures is that a
Planck function fits the data well for the portion of the spectrum
below $kT$, i.e. at low energies, which is approaching the
Rayleigh-Jeans spectrum. Also at the high energies all but
triggers 6504 and 7295 fit the data acceptably well. These two
triggers have additional high-energy flux compared to the best-fit
black-body model and there is clearly need for an additional,
weaker spectral component that affects the spectrum at high
energies. They will therefore be further studied below. In the
figures, the residual patterns for the time-resolved fits, i.e.,
the differences between the data and the best-fit-model fluxes in
units of $\sigma$, are shown below every spectrum. They are small
and do not show any strong systematic trends, something that would
appear in poor fits. Therefore, we conclude that the five triggers
in the sample are consistent with the picture of them being
dominated by emission from an optically thick environment, or a
photosphere, throughout the evolution during the pulses.

To further illustrate the modelling, the time-resolved values of
$\chi_\nu ^2$ are plotted as a function of time for all five cases
in Figure \ref{fig:chi}. These $\chi^2_\nu$-values are all
reasonable and they do not show any strong temporal evolution,
disregarding trigger 6630 (which is not  one of the thermal
pulses), which is indicative of good fits. In addition, the
$\chi^2 _\nu$ for the whole pulse is shown in the Figure as the
dashed line, for each case. This is the summed $\chi^2$-value for
all the spectra divided by the total number of degrees of freedom
(d.o.f.). They too have reasonable values, which is to be expected
as the time resolved fits were good.

In interpreting the observed spectra and the fits to them, it
should be noted that the low-energy bins (as well as the high
energy ones) have a lower significance compared to the bins
between, say 50 and 1000 keV, due to the detector sensitivity and
the background. The spectra in Figures \ref{fig:2193} to
\ref{fig:7295} were rebinned to increase the clarity.  The flux
error-bars are placed at an averaged energy of the new, broader
energy bin, weighted by the observed flux. This is necessary to
correctly reproduce the statistics of the observations, especially
for the high-energy bins. To illustrate this further, we plot in
Figure \ref{fig:background} the raw detection of the burst 7295.
The background count-rate is shown by the gray curve and the total
flux is shown by the solid line. The background is subtracted from
these raw data to find the burst spectrum, which is then modelled.
Especially for the cases studied here, which mainly have spectra
that fall off exponentially toward higher energies, the burst
signal quickly gets weaker, forcing the high-energy bins to be
broad to achieve a higher signal-to-noise ratio. Note, however,
that the $\chi^2$ fitting of the model is made with the full
resolution of the detector, which can be seen in the residual
plots below the spectra, in the previous figures.

\subsubsection{Is a Planck spectrum a unique interpretation?}

This fact, discussed at the end of the previous section, together
with the proximity of the peak of the spectrum to the low-energy
band-limit of the detector, makes it problematic to prove, beyond
any doubt, the thermal character of the spectra. Ideally, one
would like to reach the asymptotic photon index $\alpha = +1$
(compare Fig. \ref{fig:planck}) to demonstrate the black body
character conclusively. Several other emission mechanism do
produce hard-$\alpha$ spectra. This is the case, for instance, for
optically-thin synchrotron radiation with a small pitch-angle
distribution ($\NE \propto E^0$) \citep{LP00}, or similarly,
jitter radiation, where the magnetic fields have a correlation
length less than the Larmour radius of the electrons. Also a
typical optically-thin synchrotron spectrum ($\propto E^{-2/3}$)
can be deformed due to scattering in a hot cloud it traverses and
reach $\alpha \sim 0$ for high optical depths \citep{DB00}. The
high-energy portion of the spectrum would be largely unaffected
due to the decline in the scattering cross-section with energy.
But all these spectra have a high-energy power-law distribution,
in contrast to the exponential cut-offs observed above. Hard
low-energy spectra, combined with an exponential cut-off, are
given, for instance, by saturated Comptonization, with its Wien
peak $\NE (E) \propto E^2 exp(-E/kT)$ (see, e.g.,
\citet{liang97}). The low-energy spectra are, however, even harder
spectra than the Planck function ($\alpha = + 2$). Also the
Compton drag model \citep{ghis00} produces a hard low-energy spectrum
($\alpha \sim 1$) and a steep high-energy tail. These spectra are, in general,
broader than a Planck spectrum and the possibility, within the
model, of producing several pulses following immediately after
each other is not clear \citep{ghirlanda}. Furthermore, inverse
Compton scattering of soft, synchrotron, seed photons, by the
emitting electrons themselves (synchrotron self-Compton), will
have a spectrum that has $\alpha = 0$ at low-energy, if the
electron distribution is near to mono-energetic and the seed
photons have a self-absorbed spectrum (\citet{RL}). A physical
scenario for this to arise is discussed in detail by \citet{SP04}.
These spectra are however somewhat broad and require high
peak-energies for the $\alpha=0$ power-law to be noticeable in the
BATSE band. A broader electron distribution and/or seed photon
spectrum will also, by necessity, make the spectrum softer. Yet
another possibility, for such a hard spectrum, is optically-thin
thermal synchrotron emission. The spectra are approximately
described by \citep{P81,WZ}
\begin{equation}
\NE (E) \propto \exp \left[ -\left( \frac{4.5}{\sin \vartheta}
\frac{\nu}{\nu_c \Theta^2}  \right)^{1/3}\right]
\end{equation}
where $\nu_c = e B/2\pi m_e c$, and $\Theta = kT/m_e c^2$, with
the electron charge, $e$, the electron mass, $m_e$, the magnetic
field strength, $B$, and the angle of the emission with respect to
the magnetic field, $\vartheta$. The spectrum is thus a constant
as a function of energy (the energy flux $F_E \propto E^1$), with
a high-energy, exponential cut-off, which though is broad due to
the $1/3$ power in the exponent. The spectral distribution of a
few of these emission mechanisms are plotted in Figure
\ref{fig:planck} as an illustration.

Fitting the time-resolved spectra with a GRB-function
\citep{band93}, the thermal characters of the spectra will be
revealed by a large $\beta$ and $\alpha \sim 1$, which
approximately would reproduce the Planck function. To illustrate
this, we use a few time bins from trigger 2193 (fig.
\ref{fig:2193}). For instance, time bins 7 and 42 have $\alpha =
0.92 \pm 0.35$, $\beta=-4\pm 3$ and $\alpha = 0.8 \pm 0.6$,
$\beta=-5\pm 9$, respectively. The large uncertainties on these
parameters originate from the large uncertainties on the spectral
data points themselves. The measured $\alpha$-values of these fits
illustrate that a model predicting $\alpha=0$ cannot be completely
rejected. On the other hand, several time-resolved spectra of this
particular case have even harder spectra. For instance, time bin
30  has $\alpha =  1.5 \pm 0.6$, which is among the hardest
spectra yet to be reported, beside for GRB 910927 (829) with
$\alpha = 1.6 \pm 0.3$ (\citet{crider}; see also
\citet{ghirlanda}), and which is indeed consistent with a Wien
spectrum.

In summary, the hard $\alpha$-values, combined with the relatively
narrow spectrum with the exponential, high-energy cut-off, is
therefore a strong indication of the spectra actually being
black-bodies.

\subsubsection{Evolution of $kT$}

The second conclusion that is apparent from these fits is that
there is a strong spectral evolution, in that the temperature of
the black body evolves. In Figure \ref{fig:BBkT} the temperature,
or $kT$ (in the observer frame, $kT = kT' \,  2 \Gamma$, where
$\Gamma$ is the bulk Lorentz factor), of the five bursts are shown
as a function of time. The first observation to be made is that
all the pulses are similar, in particular in that the temperature
decreases monotonically even during the initial pulse rise. Such
pulses are known as hard-to-soft pulses. It is also obvious from
the plots that a change in the temperature behavior occurs at
approximately the time of the flux peak; a slower decay turns into
a more rapid one. We also note that a modification to the pure
monotonical decrease exists in triggers 4157 and 7295. These
features, in $kT(t)$, are connected to the secondary pulses in the
light curves, around 7s and 10s, respectively.

To be able to model the observed time-evolution of the temperature
we derive an analytical model for a broken power-law following
Ryde (1999). We want a function describing  a smoothly broken
power-law, having a low-energy power-law with index $a$, and a
high-energy power-law with spectral index $b$. The duration of the
transition should be a variable parameter, allowing for
differently sharp transitions. We therefore define the following
properties: its logarithmic derivative varies smoothly from $a$ to
$b$, with the transition described by a hyperbolic tangent
function. \noindent
\begin{equation}
\frac{d \log kT}{d \log t} = \xi \tanh \left(\frac{\log(t/t_{\rm
0})}{\delta}\right) + \phi, \label{eq:smooth1}
\end{equation}
where $t_{\rm 0}$ is the time of the break, $\delta$ is the width
of the transition, $\xi = (b - a)/2$, and $\phi = (b + a)/2$.
Integrating equation (\ref{eq:smooth1}) gives
\begin{equation}
kT (t) = kT_{\rm n} \left( \frac{t}{t_{\rm n}} \right) ^{\phi}
\left[ \frac {{\rm cosh}\left(\frac{{\rm log}(t/t_{\rm
0})}{\delta}\right)} {{\rm cosh}\left(\frac{{\rm log}(t_{\rm
n}/t_{\rm 0})}{\delta}\right)} \right] ^{\xi \delta {\rm ln}10},
\label{eq:smooth2}
\end{equation}
\noindent which is the function that describes the flux as a
broken power law in time, and that will be used in fitting the
data. The function is normalized at time $t_{\rm n}$ at which the
temperature is $kT_{\rm n}$, and the width of the transition, in
linear time, is $\Delta t = t_{\rm 0}(10^\delta - 1)$.  The
function is described by four parameters apart from the
normalization, namely, the two power law indices, $a$ and $b$, the
break time, $t_{\rm 0}$, and the time interval over which the
temperature changes from one power law to the other, $\delta$ or
$\Delta t$.

The results of the fits are summarized in Table \ref{tab:Tt}  and
the fits are shown as solid and dashed lines in Figure
\ref{fig:BBkT}. The data are well fitted by the smoothly broken
power-law [eq.(\ref{eq:smooth2})]. The best constrained burst is
trigger 6504, which interestingly has $b = -0.67 \pm 0.07$,
consistent with $-2/3$ and has a curvature parameter $\delta \sim
0.15$. For most of the cases, however, a fit with all the
parameters free cannot be constrained. In particular, the
$\delta$-parameter is not constrained at all. Its value is
therefore chosen to be frozen to $\delta = 0.15$ (taken from
trigger 6504) as a first fit attempt. These fits give the result
that burst 3256 also has a $b$-value close to $-2/3$
($-0.69\pm0.04$), while triggers 2193 and 7295 have a slightly
larger value, closer to $-3/4$ ($-0.78\pm0.04$ and $-0.74 \pm
0.04$, respectively). There could thus be two different preferred
decay behaviors. Such a conclusion has, naturally, to be tested on
a larger sample of bursts to have any validity. To examine whether
the latter two pulses are indeed consistent with $b = -2/3$, they
were refitted with the $b$ parameter fixed instead, and letting
$\delta$ be free to vary. These fits are also presented in the
table and the figure, where they are shown as dashed lines.
Statistically, these fits are in principle as good as the fits
found with the $\delta$ parameter frozen instead of $b$. The $R^2$
changes from $0.975$ to $0.971$ and $0.99137$ to $0.99023$ for
2193 and 7295, respectively. The main difference in the fit
parameters is that the break times, $t_0$, become smaller. We note
that these times are more consistent with the position of the peak
in the light curve in Table \ref{tab:one}. It is a reasonable
assumption that the change in temperature decay should coincide
with the peak of the light curve, which makes these fits yet more
reasonable. Furthermore, the imposed constraint above, that
$\delta$ should be the same and constant for all bursts, is not
well motivated. On the contrary, the breaks in the temperature
decay appear to be sharper, and all are actually consistent with
$\delta \sim 0$. It can also be noted that the early-time
power-law becomes more or less constant for trigger 7295. The
post-break $kT$ evolutions are thus all consistent with a $-2/3$
decay. More data on similar pulses is required to confirm or
refute that $b= -2/3$ is a universal number for all thermal
pulses.

As mentioned above, trigger 4157 has a more complicated
temperature behavior (Fig. \ref{fig:BBkT}c). It could however
easily be interpreted as both pulses showing a separate broken
power-law, resembling the previous cases, and the total
temperature curve having two breaks at $t \sim 1$ and $\sim 8$ s,
coinciding with the peaks in the light curve. We therefore fit the
data with two smoothly-broken power-laws added to each other. To
reduce the number of free parameters, we assume the $a$, $b$, and
$\delta$ parameters to be the same for both. The best fits have a
positive $a = 0.4 \pm 0.2$ and $b = -0.8 \pm 0.1$ and a very sharp
$\delta = 0.0 \pm 0.1$  ($\chi_\nu ^2 = 1.09$). Again the break is
required to be very sharp. However, $b$ is consistent with both
$-2/3$ and $-3/4$. A similar problem, with a secondary pulse,
arises for trigger 7295. Here the time bins 8 -- 12 are excluded
from the fits, thus eliminating the effect of the secondary pulse.

In conclusion, we find that  temperature decays in these thermal
pulses are well described by a smoothly-broken power-law and that
the early-time power-law is either flat or weakly decaying with an
index down to $a = -0.25$, while for the late-time power-law
index, $b$ all are intriguingly close to the value $-2/3$. It is
interesting to note that the latter is what is expected from an
adiabatically expanding fireball (see more \S
\ref{sec:interpretation}).

\subsection{Extra Spectral Components}
\label{sec:sp}

\subsubsection{Black-Body Cases}

As mentioned above, two of the five thermal pulses have an
indication that there is a weak, additional, high-energy
flux-component. In trigger 7295 (Fig. \ref{fig:7295}) the
Rayleigh-Jeans portion of the spectrum fits the data well, but
above $\sim$ 700 keV the data do not follow the exponential cutoff
of the Wien portion. The photon flux remains constant at an
approximate 5 \% level of the maximum. Similarly, but somewhat
less pronounced, there is extra flux at high energies in trigger
6504 (Fig. {\ref{fig:6504}), mainly for the earliest time-bins.
This could be due to Comptonization of the black body photons by
the hot electrons, or it could be due to a separate non-thermal
emission from synchrotron cooling of the electrons. We therefore
re-fit these spectra with a two-component spectrum: a black body,
representing the thermal emission, combined with a single
power-law representing the non-thermal emission:
\begin{equation}
N_E (E,t) = A(t) \, (E/E_{\rm piv})^s
\end{equation}
where $A$ is the amplitude in photons s$^{-1}$ cm$^{-2}$
keV$^{-1}$ and $E_{\rm piv}$ is a pivot energy in keV, and finally
$s$ is the power-law index. We assume that the non-thermal
emission, within the observed energy-band, can be represented by
this power-law. Ideally, more elaborate models should be used for
the non-thermal emission, such as a broken power law, with a peak
representing the emission from the electrons with the lowest
Lorentz factors. However, this is not permitted by the quality of
the data, and therefore the power-law index represents an averaged
value. On the other hand, we also note that for very efficient
cooling the spectrum can be expected to be a power law of $\alpha
= -3/2$ over the whole energy window.  This is especially the case
if there is no continuous heating during the emission episode. In
Figure \ref{fig:extra} one time bin for each of these two bursts
are shown with fits to such a two-component model. Note that the
spectra in the figure are in an $E \, F_E$ representation, i.e.,
energy flux per decade (unit: keV cm$^{-2}$ s$^{-1}$), to enhance
the spectral energies with most energy flux output. From the
figures it is evident that the two-component model can account for
the high-energy flux as well, which a pure black body had problems
with. The indices of the power-law component, for these fits, vary
for 6504 from $s = -0.5$ to $-1.5$ and for 7295 it is constantly
around  $s = -0.9$, with somewhat large errors.

\subsubsection{Bursts that are Initially Hard}

Several authors have pointed out bursts which have hard spectra,
consistent with black-body emission, for a short initial phase of
the light curves, but which later become non-thermal. In Figure
\ref{fig:lc2} the light curves of four such cases are shown and
the phases during which the spectra are well fitted by black
bodies are indicated. They are all medium-complex bursts. Trigger
6630 however looks like a single pulse, but it cannot be fitted
with the pulse model by \citet{KRL}, which was shown to be able to
fit most single pulses. This suggests that 6630 actually consists
of several pulses which are heavily overlapped.

Inspired by the success of the two-component model in the study of
the black-body pulses above, we continue by modelling these four
triggers with such a model as well. The results of the analysis on
trigger 5773 is shown in Figure \ref{fig:5773_4}. Only the first
20 seconds of the burst's HERB data were recorded by BATSE,
preventing us from a longer study. However, the main features are
already present in these 20 s (which include the major part of the
burst.) Panel a) in Figure \ref{fig:5773_4} shows the time
evolution of the $\chi^2_\nu$ -values for the two models: a pure
black body and a combination of a black body and a power-law. For
the first 5 seconds the black-body model gives reasonable
$\chi^2_\nu$-values, after which the fit rapidly deteriorates.
However, the two-component model is able to describe the data
throughout. The temperature of the black body, in the
two-component model, is depicted in panel b) where also the light
curve is shown by the grey line. During the initial phase the
temperature decreases monotonically, but rises again during the
first major pulse. This is repeated for the second pulse. We note
that the peak temperature precedes the peak in the light curve.
The maximum temperature of the peaks also decreases, even though
the intensity of the peaks increases. In panels c) and d) the
evolution of the power-law component is shown. The power-law index
varies from $\sim $ -0.7 down to $\sim$ -2.2. Assuming an
optically-thin cooling model and that the single power-law
averages such a spectrum, the first phase of the burst would be
dominated by the spectrum below $E_{\rm p}$ having the value
$-2/3$ (dashed line in the Figure). During the transition, which
occurs at $\sim$ 10 seconds (i.e. during or somewhat after the
first pulse), the smooth break in the non-thermal spectrum
dominates in the observed band. Later, the synchrotron spectrum
above $E_{\rm p}$ will dominate. In the fast cooling regime, the
index of the power law of the electron distribution will be
$-(p+2)/2$, where $p$ is the injected index, which for the Fermi
type of particle acceleration in relativistic shocks is $p \sim
2.2$. This value gives a photon index of $s = -(p+2)/2 = - 2.1$
(dashed line in the Figure). Finally, panel d) shows that the
amplitude of the power law grows in importance relative to the
total emission, indicating again that the thermal emission
component becomes increasingly swamped by the non-thermal
radiation.

A similar analysis was performed on trigger 6630.
\citet{ghirlanda} demonstrated that the first $\sim$ 2 s are
consistent with a black body and that the fits get worse
thereafter. Here again the two-component model can indeed explain
the data and the fits are good throughout the burst. To illustrate
this we plot, in Figure \ref{fig:6630}, two time-resolved spectra
which are beyond the initial black-body phase. Note again that the
spectra in Figure \ref{fig:6630} are in an $E \,F_E$
representation. These spectra are poorly fitted by a model
consisting of a pure black body, which can be seen in Fig. 9 in
\citet{ghirlanda}, where both the $\chi^2$ as well as the
residuals reflect the poor fit. However, as can be seen by the
best fit for the two-component model, which is shown in Figure
\ref{fig:6630} by solid lines, such a model can describe the data.
The $\chi^2_\nu$ for the two fits are now $\chi^2_\nu (dof) = 0.81
(105)$ and $ 0.995 (105)$, respectively. Note, in particular, the
improvement in the systematics of the residuals for the spectrum
of time bin $t=2.75 - 3.1$ s, compared to the pure thermal fit in
\citet{ghirlanda}. Throughout the fit, the power-law index varies
from $s = -1.1$ down to $s = -2.4$, suggesting that the power-law
component is a synchrotron spectrum from the electrons above their
minimum Lorentz factor.

A two-component model is also satisfactory for the initially hard
triggers 829 and 1085, giving $\chi^2_\nu$ and residuals that are
acceptable. The power-law indices vary from $s \sim -1.5$ to $\sim
-2$, and $s \sim -1$ to  $\sim -2$, respectively. We also note
that \citet{ghirlanda} showed, for a few of their triggers, which
were initially hard, that the early-time temperature decreases as
$kT \propto t ^{-1/4}$, consistent with the early power-law decays
of the thermal pulses in the previous section. Finally, inspection
of the light curves in Figure \ref{fig:lc2} suggests that the
initial black-body behavior could be connected to a separate
pulse. This is consistent with the picture that it represents a
photospheric flash and the later, overlaid pulse(s) are from the
non-thermal emission region. The analysis shows also that the
photospheric temperature rises just before subsequent pulses,
representing several different photospheres.

The interplay of the strengths of these two components will
obviously affect the measured low-energy power-law of the
spectrum, if it is modelled by a single GRB-function
(eq.[\ref{eq:band}]). The variation of the component amplitudes
will manifest itself as an evolution of the $\alpha$-parameter.
Such a behavior has indeed been observed to occur in many bursts.
This evolution therefore might simply reflect the relative
strengths between the thermal and non-thermal components, and does
not necessarily need to represent an actual change in radiative
regime where the single electron emission pattern has to change,
such as a change from self-absorbed to optically-thin synchrotron
emission from the same emitting material. As the spectra in
general are observed to soften with time, it would indicate that a
thermal-dominated phase is followed by an increasing importance of
the non-thermal component.

Finally, we note that the two-component model has four parameter,
namely the black-body normalization, its temperature, and the
power-law index and its normalization. It therefore has equal
number of parameters as the GRB-function (Eq. \ref{eq:band}).

\section{Physical Interpretation}
\label{sec:interpretation}

As suggested by the observations above, several spectral
components probably are involved in the creation of the observed
gamma-ray spectrum. Suggestions for the origin of the thermal
component have been given both in the kinetic, fireball model and
for electromagnetic models. To characterize the outflow we define
the parameter $\sigma = L_{\rm P}/L_{\rm kin}$ as the ratio
between the magnetic energy flux, or Poynting flux, and the
kinetic energy flux. We first discuss our results within the
fireball model (small $\sigma$) and later expand the discussion to
also include large-$\sigma$ winds.

\subsection{Fireball Model, small {$\sigma$}}
\label{sec:fireball}

In the fireball model the GRB emission stems from  a highly
relativistic wind with a low baryonic load. Large amounts of
energy is assumed to be released at a small initial distance from
the progenitor, $R_0$, and are injected into an optically thick,
electron-positron  pair wind (or shell). Several physical
settings, where this could occur are plausible, but for long GRBs,
it is now generally assumed that this occurs close to the center
of a collapsing massive star. The wind is radiation-dominated and
accelerates under its own pressure as a relativistic gas and, as
it is opaque, the main part of the radiative energy is initially
trapped. The opacity stems from the ambient baryonic electrons and
the e$^{\pm}$ pairs (the pairs naturally make a contribution only
if they have not annihilated). During the expansion the fireball
cools adiabatically, converting its internal energy into kinetic
energy. Its Lorentz factor increases linearly \citep{MLR} until it
reaches the saturation radius, $R_{\rm s} \sim R_0 \eta$, where
the dimensionless entropy of the wind is given by $\eta =
L/\dot{M}c^{2}$ and describes the baryon load of the fireball (the
initial radiation to rest-mass-energy ratio). Beyond this point
all the thermal energy has been converted into kinetic energy of
the out-flowing baryons, and the fireball coasts with constant
Lorentz factor, $\Gamma = \eta$. The emission from this fireball
is expected to be thermal. The opacity decreases while the
fireball expands and the main part of the thermal radiation is
emitted when the optical depth, $\tau$, is around unity.  The
transition time to optical thinness is relatively short, after
which the emission becomes optically thin.

We will discuss two variations of the fireball model. One can
imagine a continuous relativistic wind or one can focus on a
single optically-thick shell and follow its emission as its
expands.

\subsubsection{Wind Model} \label{sec:model1}

A typical scenario in which a photospheric pulse can be created is
the one described by a relativistic wind \citep{bp}. The wind will
have a photosphere at a radius and a temperature that are only
dependent on the initial parameters of the flow. If one assumes a
constant flow, the photospheric emission will be time-independent.
For it to give rise to a pulse shape, the energy injection rate
and/or the Lorentz factor have to change.

The wind photosphere is defined as the surface at a distance
$R_{\rm w,ph}$ where $\tau_{\rm w} = 1$. Above the photosphere the
wind is optically thin. Let us calculate the optical depth through
the whole wind measured along a ray towards the observer at
infinity:
\begin{equation}
\tau(R_{\rm ph}) = \int_{R_{\rm ph}} ^{\infty} \, d\tau (R)
 = \int _{R_{\rm ph}} ^\infty \frac{\rho' \kappa
dR}{{\cal{D}}(\mu)}. \label{eq:int}
\end{equation}
$\kappa$ is the total mass opacity (per gram) due to the ambient
electrons and the $e^{\pm}$ pairs, and the comoving density is
given by
\begin{equation}
\rho' = \frac{1}{\Gamma} \frac{\dot{M}}{4 \pi R^2 \beta c}.
\label{eq:rho}
\end{equation}
 Note that $\beta \Gamma \sim \Gamma$ for large
$\Gamma$. The angle-dependent Lorentz factor ${\cal{D}}(\mu)
\equiv [{\Gamma(1-\beta \mu)}]^{-1}$, takes care of the angular
dependence of the optical depth \citep{abram} in which $\theta =
\rm{arccos} (\mu)$ is the angle between the direction of photon
propagation and the matter velocity, which for small angles
reduces to ${\cal{D}} = 2 \Gamma$. The upper limit in equation
(\ref{eq:int}) represents integration through the whole wind.
Solving this equation for $\tau = 1$ gives that the wind
photospheric radius is at
\begin{equation}
R_{\rm ph} = \frac{\kappa \dot{M} }{4 \pi \beta c \Gamma
{\cal{D}}(\mu)} \sim \frac{\kappa \dot{M}}{ 8 \pi c \Gamma^2}
\label{eq:Rphot}
\end{equation}
In the last step we used the fact that a typical burst $\Gamma$ is
large and that $v \rightarrow c$. If the mass-injection rate is
constant, then $R_{\rm ph}$ will remain constant and have a
certain temperature, $kT(R_{\rm w,ph})$ and luminosity, $L_{\rm
ph} \propto \dot{E}$. The temperature of the adiabatic,
relativistic outflow depends on the radius as $kT(R) \propto
R^{-2/3}$ (standard fireball model, \citet{piran99}). To account
for the temporal shape of a pulse one therefore has to assume a
certain density profile in time of the wind or certain variations
in the $\Gamma$ and/or $\dot{M}$. The location of the photosphere,
and thereby the luminosity and temperature, has to be determined
numerically. The observed shape of a pulse can then be used to
deduce the history of the wind. In the pulse model of \citet{DM},
the rise phase of a pulse was attributed to a varying Lorentz
factor. After the initial rise, the Lorentz factor remains
constant and therefore also the intensity, until the end of the
wind.
Overlaid will be the non-thermal emission from the internal
shocks, which together create the observed pulses.

To model a pure thermal pulse, including the decay phase, any of
the other parameters have to be changed, for instance $\dot{M}$
can be varied. Then $\dot{M}$ has to grow approximately linearly
to explain the observations above in Figure \ref{fig:BBkT}. The
wind model also requires that $kT(R) \propto F(R) \propto
R^{-2/3}$, that is, the ratio between temperature and flux,
$kT/F$, is constant. A temporal variation in $kT$ will be followed
by a similar variation in the flux\footnote{In the language of
\citet{RP02} the HIC $\eta = 1$}. However, an increase of the
temperature during the rise phase of the studied pulses above is
not observed (see also \citet{ghirlanda}). A resolution to this
could perhaps be the curvature effect. If it indeed is dominating,
a monotonic flux-decay  in the comoving frame (and thus a
monotonic temperature decay), would have a pulse shape in the
observer frame, see for instance figure 6 in \citet{RP02}.

It is also plausible that the optical depths, and thus $R_{\rm
ph}$ and $kT$, are different for different energies, which would
lead to a multi-color black-body, decreasing the thermal
appearance of the observed spectrum. On the other hand,
\citet{RS00} noted that several pulses have a break in their
power-law decays, with the decay turning to a faster fall-off. In
the wind model of \citet{DM} such features are naturally explained
and in the light curves that they produce, these features are
indeed present, see their Fig. 6. Competing models therefore would
have to explain these breaks as well. Here, we see such behaviors
in the pulse of trigger 829, at 14 s, and in the pulse of 6630, at
5.5 s. These pulses are shown in Figure \ref{fig:lc2}, in which
the breaks are identified by the arrows.

In summary, the wind model does not provide an obvious reason for
why the observed, fast temperature decays all seem to be narrowly
distributed and tend to have a decay index of approximately $\sim
-2/3$. However, more numerical investigations will be necessary.

\subsubsection{Shell Model} \label{sec:model2}

We now study the following scenario. A hot, geometrically thin,
fireball shell, with large optical-depth is ejected from the
progenitor source into a low-intensity wind. The photosphere of
the {\it wind} should be at small radii to ensure that the thermal
{\it shell} emission occurs in an optical-thin environment of the
wind, in order that it should be visible to an observer at
infinity. From equation (\ref{eq:Rphot}), $R_{\rm ph}$ is small
if, for instance $\dot{M}$ is small and/or $\Gamma$ is large. The
shell can still be optically thick, $\Delta \tau > 1$, at the wind
photosphere. For a shell with width, $\Delta R$, the optical depth
is $\Delta \tau = (\kappa \dot{M}_{\rm sh} \Delta R)/(8\pi R_{\rm
ph,w}^2 \Gamma ^2 c).$ Moreover, we note that as long as the
variations of the $\Gamma$ internally to the shell is not too
large in relation to the width of the shell, it will not expand
significantly from $\Delta R$ and behave as a section of a wind.

We now assume that the rise phase of the pulse can be attributed
to the acceleration phase of the fireball, i.e. $R < R_{\rm s}$,
with the energy flux $F \propto \Gamma^k$(t), to some power $k$,
as well as to the effective emitting area. As the plasma is
optically thick to its own electrons it is in principle an
isentropic fluid which can be modelled as a gas with adiabatic
index $\gamma = 4/3$ \citep{RL}. We therefore have
\begin{equation}
kT' \propto {\rho'} ^{\gamma - 1} = {\rho'} ^{1/3}
\label{eq:adiab}
\end{equation}
which is the adiabatic relation for electromagnetic radiation.
Assuming a power law increase of the Lorentz factor, $\Gamma
\propto t^{\xi}$
\begin{equation}
\rho' \propto R^{-2} \Gamma^{-1} \propto t^{-(2 + \xi)}
\end{equation}
so that
\begin{equation}
kT = kT' \, \Gamma \propto t^{\, 2/3 \, (\xi -1)} \label{eq:kta}
\end{equation}
If $\xi = 1$, then the observed $kT$ is constant. The fits to the
cases in Figure \ref{fig:BBkT} indicate that $\xi$ has values from
$0.6$ to $1.6$, see Table \ref{tab:Tt}. In the numerical study by
\citet{MLR}, it was found  that $\Gamma$ initially increases
linearly at early stages of the evolution and later saturates (see
their Fig. 3).  The data here seem mainly to prefer a slower
acceleration (we will later see that large-$\sigma$ models do
indeed require such an acceleration). It is interesting to note
that in the numerical modelling by \citet{remo} (and references
therein) the Lorentz factor increases linearly at a beginning but
turns to a slower than linear increase after the pair fireball has
interacted with the baryonic remnant of the progenitor star,
before it saturates (their Fig. 9). The break-width parameter of
the fitting function used here, $\delta$ (Eq. [\ref{eq:smooth2}])
is of interest as it describes the transition from the linear
increase of $\Gamma$ during the acceleration phase, to the
saturated, coasting regime.

The break in the temperature decay is now reached at the
saturation radius before the shell becomes optically thin.
The photosphere of the fireball shell should then be at large
distances from the progenitor to ensure that the emission is
thermal throughout the whole evolution. From equation
(\ref{eq:Rphot}) we have that $R_{\tau=1} = 10^{15}$ cm for  large
$R_0$ and low $\eta$ values. $10^{16}$ cm is a typical radius at
which significant deceleration of the fireball has occurred as a
result of the interaction with the interstellar medium and the
onset of the afterglow (see, e.g. \citet{ree92}). The
optically-thick shell is cooling adiabatically due to the
expansion and during the coasting phase ($\Gamma$ constant) the
cooling of the shell will cause the intensity to drop and thus
produce the decay phase of the pulse. The temperature changes as
$kT' \propto {\rho'}^{1/3}$ (eq. [\ref{eq:adiab}]) and if the
density of the wind decreases as $\rho \propto R ^{-2}$ (eq.
[\ref{eq:rho}]), the adiabatic relation gives $kT \propto R
^{-2/3}$. As $\Gamma$ is constant,
\begin{equation}
kT (t)  \propto t^{-2/3}. \label{eq:kTadiab}
\end{equation}
which corresponds to the observed temperature decays beyond the
break.

A suggestion of such a scenario is given in \citet{BSM}, who
investigated the evolution of a pair fireball before and after it
becomes optically thin. The collision-dominated pair phase results
in the flash of thermal emission that is followed by the
transition phase during which the fireball becomes Thomson thin,
but the radiation remains dominated by thermal Comptonization. The
initial thermal flash should typically  be seen by BATSE, and
could even be the main component. The flash from the optically
thinning process would then be seen at higher energies, between
100 MeV and a few GeV, that is, much beyond the BATSE window.
They find that the time for the fireball to become Thomson thin
can be extended out to $\sim 1$ s and beyond depending on the
total energy and the collimation (see their Fig. 1). The main
characteristics of the observed pulses, can therefore, in
principle, be reproduced.

An alternative to this scenario is, of course, that the thermal
flash, emitted as the shell becomes optically thin, is detected in
the BATSE range, instead. The break is thus reached during the
acceleration phase as the transparency condition is met.
\citet{BSM} showed that up to a few seconds after the initial
explosion the pair plasma is still dominated by collisional
processes and that the collision-less shocks, that are responsible
for the non-thermal radiation,  only occur after that turbulence
has had time to form. This ensures that thermal character of the
spectra. \citet{grim} also showed that for a relativistic e$^\pm$
pair wind the radiation field will be approximately black body out
to $\tau << 1$ as long as $\Gamma \propto R$. If the optically
thinning transition phase is reached  as the fireball shell is
still accelerating, the constancy in temperature during the rise
phase can be explained, according to the above, and the decay is
due to the interplay between the ongoing acceleration and the
radiative emission, and detailed numerical simulations are needed,
which again need to be able to explain the alleged preference of
the -2/3--temperature drop. Furthermore, the modelling has to
produce the peak energy at lower energies ($\sim$ 200 keV).

In summary, an optically-thick shell, expanding adiabatically, can
reproduce the observed behavior, with the break in the temperature
decay occurring at the saturation radius.

\subsubsection{Additional Spectral Components and Additional Pulses}

The above considerations were mainly for a purely thermal, single
pulse. However, for a few of the investigated pulses there is an
indication of a weaker, secondary, non-thermal spectral component
as well as secondary pulses. We note again that the temperature
rises in advance of the new pulse. Also, the non-thermal component
becomes most important after the peak.

The typical photospheric radius is well before the radius at which
the internal shocks occur $R \sim 3 \times 10^{14} {\rm cm} \;
(\Gamma/100)^2$, which is somewhat dependent on the initial
distribution of the Lorentz factors. If the collisions occur
mainly below the photosphere the outflow is more ordered in the
optically-thin region and thus the shocks are less efficient,
since the relative Lorentz factors are smaller. This could explain
the absence of a non-thermal component in some pulses. Indeed,
\citet{KRM} showed that different amounts of variability of the
shock emission can easily be found for varying initial set-ups.
The non-thermal emission naturally arises in a time period after
the thermal emission escapes, due to the time scale needed for
turbulence to be created \citep{BSM}, thereby explaining the
delay. The initial section of a burst should bear the strongest
signal of the photosphere.

Subsequent pulses could be due to the release of additional,
optically thick shells from the central engine repeating the
behavior. If the central engine emits another major shell which
catches up the first one, they will collide and merge to
temporarily form a single shell. The internal energy of the merged
shell is the difference in kinetic energy before and after the
collision. A significant fraction of the kinetic energy can
therefore be converted into internal energy if the relative
velocity is relativistic. The collision will thus re-energize the
merged shell, heating it to higher temperatures and giving rise to
the new thermal pulse in the light curve. A while later,
synchrotron radiation is emitted due to the turbulence excited by
the collision in the shell photosphere.

\subsection{Poynting-Flux Models, Large $\sigma$}

Most progenitor models for GRBs, include the formation of a
rapidly rotating, compact object. Indeed, in the collapsar model
in which the GRB is emitted when the iron core of a massive star
collapses can very well produce a very rapidly rotating, highly
magnetized, neutron star \citep{WZ}. The magnetic field on such an
object can be as high as $10^{16}$ G. The electromagnetic torque
that such a field exerts on the compact object will partly
decelerate its rotation on a time scale of seconds, and partly
generate a strongly magnetized wind flowing at high Lorentz
factors (see \citet{uzov} for a review). Such magnetic winds
therefore naturally arise in most formation scenarios of GRBs. The
importance of such a magnetic Poynting flux, $L_{\rm P}$, compared
to the kinetic wind, $L_{\rm kin}$ (which is the luminosity in
both e$^\pm$-pairs and radiation) is given by the
$\sigma$-parameter, and $L_{\rm P}$ depends on high powers of the
radius, magnetic field and  the angular velocity of the compact
object. For compact objects with a very high magnetic field,
$\sigma$ is expected to be relatively large \citep{uzov}. Also in
electromagnetic models, the transformation of the out-flowing
energy into gamma-rays is more efficient than in the kinetic
models, and they naturally avoid heavy baryon loadings.

If the magnetic configuration of the rotating compact object is
non-axisymmetric (such as the rotation of an inclined dipole) the
resulting Poynting flux could very well be in the form of a
striped wind, that is, the magnetic field is aligned transversely
to the flow direction and changes sign on small length scales.
Reconnection then creates a natural possibility for the local
dissipation of the magnetic energy which is transferred to the
matter. Apart from releasing energy, that can be radiated away and
observed by a distant observer, this leads to a decrease in the
magnetic energy with $R$, producing an outward gradient in the
magnetic pressure, which is a source of acceleration of the flow.
The rate of dissipation of the magnetic energy depends then on the
reconnection rate of closely lying regions of different field line
directions. Below the photosphere all the energy is deposited in
the outflow and its adiabatic expansion. The acceleration here is
slower than that in the fireball scenario, due to the continuous
energy release; $\Gamma \propto R^{1/3}$ instead of the $\Gamma
\propto R$, discussed above. As mentioned in \S \ref{sec:model2}
this is indeed what the temperature decay of most of the observed
cases seem to indicate. The observed temperature of the thermal
black-body emission from the photosphere is $ kT = 2 \Gamma kT_0$,
where $T_0 \sim 2\times 10^8$ K $= 17$ keV/k. The Lorentz factor
at the photosphere then has to be $\sim 10$ to explain the
observed peak energies in the previous section, which can be found
for a somewhat less strong magnetic field. The saturation radius
for the fireball model is where the terminal bulk Lorenz factor is
reached. Such a radius also exists for a Poynting flux wind. As
the rate of dissipation decreases due to the decreasing field
strength, it eventually becomes slower than the expansion
time-scale, and the dissipation terminates. The flow therefore
ceases to accelerate and reached a final speed. After the
out-flowing plasma has become optically thin and the magnetic
field ceases to be frozen to the wind plasma (as the
Goldrich-Julian density is reached) and large-amplitude
electromagnetic-waves  are created, in which the out-flowing
particles are efficiently accelerated and emit non-thermal
radiation.

Depending on where most of the energy is dissipated relative to
the wind photosphere, the observed emission will be different, in
much the same way as in the fireball discussion in the previous
sections. In the detailed numerical modelling by \citet{DS} of a
Poynting flux wind, they illustrated the different scenarios by
varying the $\sigma$-parameter. For low values most of the energy
is converted into kinetic energy already before the photosphere.
At medium values the thermal component becomes important. At most
$\sim 17 \%$ of the dissipatable magnetic energy is released as
black-body emission. At higher $\sigma$, the main part of the
magnetic energy is dissipated in the optically-thin region, and
mainly non-thermal radiation is observed. Therefore, such a wind
will also naturally produce several components in the observed
spectrum, with variable relative strength (see also \citet{Lyu}).

\citet{DS} further showed that the main properties of the emission
essentially depend on the ratio between the saturation radius and
the photospheric radius. The black-body component will be most
important when these two radii coincide. This means that if we see
black-body emission dominating the spectrum, as in the observed
cases above, we are probably seeing outflows in which the
black-body emission is maximally important and thus the saturation
and photosphere radii are similar. In such a set-up the initial
Poynting flux is of the order of 40 times larger than the kinetic
energy. The wind accelerates with $\Gamma \propto R^{1/3}$ up to
the point where the photons escape, at the photosphere. This model
could thus overcome the problems in the previous section, in which
it was assumed that the saturation radius was reached in the
optically-thick region, which makes the emission inefficient.
Here, instead, as the saturation and the photospheric radii
coincide, the peak emission does indeed emanate from the
photosphere. Also, inserting $\xi =1/3$ in equation (\ref{eq:kta})
we obtain $kT \propto t^{-4/9}$ which is steeper than in the
fireball model and might be a better description of the early
temperature decays, most notably, trigger 2193. These features
make the Poynting flux model attractive to explain the
observations, even though a final answer will need to include more
detailed modelling, beyond the scope of this paper.

Secondary pulses of the light curve, in this model, might be due
to secondary outbursts from the central engine. If the properties
of the second outburst are different from the first one, for
instance a decreasing $\sigma$-ratio (maybe by the baryon
'pollution' increasing with time), the photosphere could be
reached at larger distances and the temperature of the thermal
radiation will then be lower, explaining this property in the
observations. Also, the non-thermal emission, that stems from the
dissipation in the optically-thin regions through reconnections
happening beyond the photosphere, will be emitted with a small
time delay compared to the thermal emission. This is what is
observed.

In summary, similarly to the shell model, this model can explain
the observed behavior. In this case the saturation radius and the
photosphere coincide. Furthermore, a slower acceleration can
reproduce the finite temperature decay slopes before the break.

\section{Discussion and Conclusions}
\label{sec:disc}

There are two important regions in the relativistic outflow that
produce the high-energy emission in GRBs: the thermal photosphere
and the region in the wind emitting non-thermal radiation. This
leads to multiple components in the light curve and spectra of
GRBs. We have identified a few simple GRB pulses which are
consistent with an intrinsic black-body emission. It is therefore
natural to interpret these exceptionally hard pulses as being
emission from the wind (or shell) photosphere. Most of these
bursts have good black-body fits throughout their evolution. In
two cases, however, (triggers 6504 and 7295), there is evidence of
a weak, non-thermal component. Furthermore, there are other,
medium complex, bursts with an initial thermal phase and a later
phase that is non-thermal. An example of such a burst is trigger
6630. We have shown that a two-component, hybrid model can indeed
describe all these bursts throughout their evolution. Several
theoretical models predict such a two-component spectrum.
Furthermore, Comptonization of the black-body component could
explain the extra flux at high energies. However, as we have
shown, the two-component model can explain the softening of
$\alpha$, which Comptonization would fail to do. We also found
that the temperature decay is a monotonic function, that can be
modelled by a broken power-law with an initially flat or weak
decay turning into a faster decay. All bursts are consistent with
a sharp break and a decay of $-2/3$.

Our interpretation is therefore the following. The spectra and
light curves of all GRBs consist of several components, a black
body and a non-thermal one with variable relative strengths. In
the five pulses in our main analysis, the black-body emission
dominates throughout the pulse. The non-thermal components are
weak or are not seen in the gamma-ray band. However, for other
cases the black-body component will be invisible or seen only in
the initial phase of the light curve.

McQuinn, Ryde, \& Petrosian (2004, in prep.) studied the curvature
effect and found that it can have a substantial effect on the
light curve, while its effect on the spectra is smaller. Our
spectral analysis presented here should therefore not be heavily
affected by this effect. They further find, as mentioned above,
that the curvature effect is more important on the light-curves of
thermal pulses compared to non-thermal pulses, even though their
spectra are not strongly altered. We therefore concentrated our
present study on the behavior of the temperature $kT$. Compared to
the photon flux (e.g. in Fig. \ref{fig:5773_4}b,d) it also has
several simplifying advantages. The measurements do not suffer as
much from additional flux components and band width problems. We
postpone the detailed analysis of the light curves, including all
important physical effect, to a later paper.

One could  argue that as the Rayleigh-Jeans portion of the
spectrum is no longer visible for some of the late-time spectra,
it might be dangerous to interpret these fits. This happens, for
instance, for the latest time-bins in triggers 3256 and 4157. As
we are not recording the spectrum beneath the energy threshold, it
may very well not be a Rayleigh-Jeans tail anymore. However, as
the study above has shown, when the spectrum starts to differ
significantly from a black body, the change will be visible mainly
at the highest energies. Thus, if the low-energy spectrum were
altered, a change in the Wien-portion of the spectrum, which is in
the observable range, would also be detected. Especially in a $E
F_E$ representation the high-energy component will show up
markedly. Therefore, we argue that the black-body fits can be
trusted even though, for some spectra, most of the Rayleigh-Jeans
portion is out of the band.

It has also been made clear that a high temporal resolution is
vital for spectroscopic studies. Due to the spectral evolution of
the spectrum during the pulse, the {\it time-integrated} spectrum
will not necessarily reflect the {\it instantaneous} spectrum,
which is the one bearing the direct consequence of the physical
creation process(es). This was shown in the comparison between the
fits in Table \ref{tab:two} and the instantaneous fits. The
connection between the instantaneous spectra and the
time-integrated spectra is, in general,  well understood and is
prescribed by the spectral, empirical relations, as discussed by
\citet{RS99}.

The study presented here has illustrated the risks of interpreting
fits made to BATSE data. In general, fits are made using the GRB
function (eq. [\ref{eq:band}]) and the resulting parameters are
then interpreted. The first issue that has to be noted is that the
low energy power-law, $\alpha$, is the {\it asymptotic} power-law
index and that the spectrum within the BATSE band seldom reaches
this power law and comprises mainly the curved section below the
peak energy. The $\alpha$--value therefore does not need to reach
$\alpha = +1$, which would be the criterium to identify a
black-body spectrum. This is the case in particular for black-body
spectra which have such a low $kT$ that the Rayleigh-Jeans part of
the spectrum is not fully detected. It is therefore risky to
directly interpret results, found by using the GRB-function, in
terms of radiation processes. As demonstrated above, the correct
approach is to fit the data directly with a Planck function (maybe
in combination with other components), which is then used in the
forward-folding, deconvolution-process, and thereafter interpret
the results of the fit. Also, as evident from the fits, an
artificial softening of the spectra would be detected, if they
were to be fitted with the GRB function. This is, for instance,
evident for trigger 4157 in Figure \ref{fig:4157}. The reason for
this effect is, again, that the black body moves out of the
detector's limited energy window. Also an apparent softening can
be caused by a changing relative strength between the hard,
black-body emission and the power-law, non-thermal emission.
Moreover, it is also important to point out that the black body is
a physical model while the GRB function is merely an empirical
model.

As illustrated  in Fig. \ref{fig:planck} the conclusive evidence
for a black body, namely the $\alpha = +1$ power-law is not
completely reached for the BATSE pulses. Further observations of
very hard spectra by, for instance,
{\it Swift}'s BAT ($\sim 15 - 150$ keV) will be able to better
prove the nature of the emission process. However, the triggering
systems of detectors are such that the observed spectra have peaks
mainly concentrated within their energy band. Therefore, a very
high peak-energy case, with a large interval, within which
$\alpha$ can be measured, will by necessity be rare. We have
further argued that a non-thermal emission component should
normally accompany the thermal one. Therefore, below a certain
energy the low-energy component will still be dominated by the
non-thermal emission. Rather than focusing on the $\alpha$-value,
a broad band detection by, for instance the {\it Gamma-Ray Large
Area Space Telescope (GLAST)}, or, even better, a combination of
{\it Swift} and {\it GLAST}, will therefore be powerful in testing
further the observations and suggestions presented here. {\it
GLAST}, which is scheduled to be launched in year 2007, will have
a full spectral coverage from approximately 10 keV up to $\sim$
200 GeV, by combining its gamma-ray bursts monitor (GBM), and its
large area telescope (LAT).  As the observations above suggest,
there are synchrotron components with peak energies higher than
$\sim $ 1 MeV in several cases. The full spectral coverage of
GLAST will give better possibilities in fitting both the thermal
and the non-thermal components and studying their relative
behaviors. Furthermore, if the pulse is indeed due to the thermal
emission from the optically-thick shell, as described above, it
must needs be accompanied by a flash at higher energies. This
would be detectable by the GLAST satellite with the combination of
the GBM, seeing the thermal, optically thick pulses and maybe the
non-thermal component, and the LAT, seeing the emission from the
optically thinning phases. Such observations will give a
possibility to test these models.

 \acknowledgments
 I am  grateful to M. B\"ottcher, G. Drenkhahn,
P. M\'esz\'aros for valuable discussions. I am also thankful to
C.-I. Bj\"ornsson, C. Fransson, V. Petrosian, J. Poutanen  and R.
Ruffini for useful comments. The anonymous referee is also
thanked. F.R. acknowledges financial support from the Swedish
Research Council. This research made use of data obtained through
the HEASARC Online Service provided by NASA's Goddard Space Flight
Center.

\newpage

\begin{deluxetable}{lllr}
 \tablecolumns{7}
 \tablewidth{0pc}
 \tablecaption{Bursts Discussed in the Text}
 \tablehead{ \colhead{Burst} & \colhead{Trigger} & \colhead{LAD} & \colhead{$t_{\rm m}$\tablenotemark{a} }
 \\
\colhead{Name} & \colhead{} &\colhead{} & \colhead{[s]} }

\startdata
 910927  &   829\tablenotemark{b}     & 4 & 3.5 \\
 911118  &   1085\tablenotemark{b}    &  4 & 6.0  \\
 930214  &   2193    &  1  &  10.7  \\
 941023  &   3256    &  7  &   1.6  \\
 951228  &   4157    &  7   &   7.8   \\
 970111  &   5773\tablenotemark{b}    & 0  & 17.2 \\
 971127  &   6504    &  2  &   3.1  \\
 980306  &   6630\tablenotemark{b} & 3 & 1.8 \\
 990102  &   7295    & 3 & 2.2  \\

 \tablenotetext{a}{Time of peak emission of the dominant pulse}
 \tablenotetext{b}{Cannot be fitted with the pulse model of \citet{KRL}}

\enddata
 \label{tab:one}
\end{deluxetable}

\begin{deluxetable}{lrrrr}
 \tablecolumns{5}
 \tablewidth{0pc}
 \tablecaption{Averaged Spectral Properties of the Burst Sample Using the GRB Model}
 \tablehead{\colhead{Trigger} & \colhead{$C$} & \colhead{$\alpha$}
 & \colhead{$\beta$} & \colhead{$E_{\rm p}$} \\
 \colhead{ } &\colhead{[photons/s/cm$^{2}$/keV]} & \colhead{} &\colhead{} &
 \colhead{[keV]} }

\startdata
 829     & $0.095 \pm 0.005$  &  $-0.03 \pm 0.05$  & $-4.0 \pm 0.2$ & $122.9 \pm 1.1 $  \\
 1085    &   $0.155 \pm 0.003$  & $-0.65 \pm  0.02$  & $-2.95 \pm  0.05$  & $209 \pm 2$  \\
 2193    &   0.0140  $\pm$ 0.0010 & 0.53  $\pm$  0.10  &      $-2.76  \pm$  0.18      & 269  $\pm$   9   \\
 3256   &   0.008 $\pm$  0.001&  $-1.03  \pm$  0.13  & $-\infty$\tablenotemark{a}   &  152  $\pm$  9  \\
 4157   &   0.11  $\pm$  0.03 &  0.10 $\pm$   0.19 &   $-3.6  \pm$  0.3  &  72.2   $\pm$  1.4   \\
 5773  & $0.131 \pm 0.002$ &  $-0.17 \pm 0.02$ & $-4.6 \pm 0.4$ & $185.1 \pm 1.1$ \\
 6504   & 0.017 $\pm$   0.003 & $-0.2 \pm$    0.2 & $-3.4 \pm$   1.4   &  179  $\pm$   14 \\
 6630   & $0.092 \pm 0.002$ & $-0.52 \pm 0.03$ & $-\infty$\tablenotemark{a}   & $283 \pm 5 $  \\
 7295    & 0.0175$\pm$  0.0013 & 0.58 $\pm$  0.12 &  $ -2.8 \pm$  0.2 &   341 $\pm$ 16
 \tablenotetext{a}{Not constrained}
\enddata
 \label{tab:two}
\end{deluxetable}

\begin{deluxetable}{lccccc}
 \tablecolumns{10}
 \tablewidth{0pc}
 \tablecaption{Temperature Evolution}

\tablehead{ \colhead{Trigger} & \colhead{$k T_{\rm n}$} & \colhead{$a$}
&\colhead{$b$} & \colhead{$\delta$} & \colhead{$t_0$}\\

\colhead{} & \colhead{[keV]} & \colhead{ } &\colhead{ } &
\colhead{ } & \colhead{[s]}

}

\startdata
 2193 & $86.4 \pm 1.2$ & $-0.25 \pm 0.02$ & $-0.78 \pm 0.04$ &$0.15$\tablenotemark{a} & $12.9 \pm 1.1$\\
      & $86.2 \pm 1.7$ & $-0.26 \pm 0.02$ & $-0.67$ &$0.01 \pm 0.15$ & $11.0 \pm 0.6$        \\
 3256 &  $17.5 \pm 0.5$ & $-0.16 \pm 0.04$ & $-0.69 \pm 0.04$ & $0.15$\tablenotemark{a} &  $2.3 \pm 0.3$    \\
 4157\tablenotemark{b} &  & &  &  &   \\
 $\;\; $ {\it comp 1} &   $6.5 \pm 1.4$ & $0.4 \pm 0.2$ & $-0.8 \pm 0.1$ & $0.0 \pm 0.1$ & $0.67 \pm 0.06$  \\
 $\;\; $ {\it comp 2} &  $13 \pm 2$ &  -"- & -"- & -"- &  $7.8 \pm 0.5$ \\
 6504 &  $28.4 \pm 0.6$ & $-0.11 \pm 0.09$ & $-0.67 \pm 0.07$ & $0.16 \pm 0.18$ & $3.8 \pm 0.6$  \\
 7295\tablenotemark{c}  &  $41.2 \pm 1.3$ & $-0.12 \pm 0.06$ & $-0.74 \pm 0.04$ &$0.15$\tablenotemark{a}  & $3.3 \pm 0.5$\\
                       &  $41.8 \pm 1.3$ & $0.04 \pm 0.21$ &$-0.67 $ &$0.2\pm 0.2$  & $2.17 \pm
                       0.7$\\
 \tablenotetext{a}{Not constrained.}
 \tablenotetext{b}{Two, added, smoothly-broken power-laws.}
 \tablenotetext{c}{The interval $t=7.2 - 13.9$ s is excluded from the fit.}
\enddata
\label{tab:Tt}
\end{deluxetable}

\newpage

\begin{figure}[]
\epsscale{0.8}
 \plotone{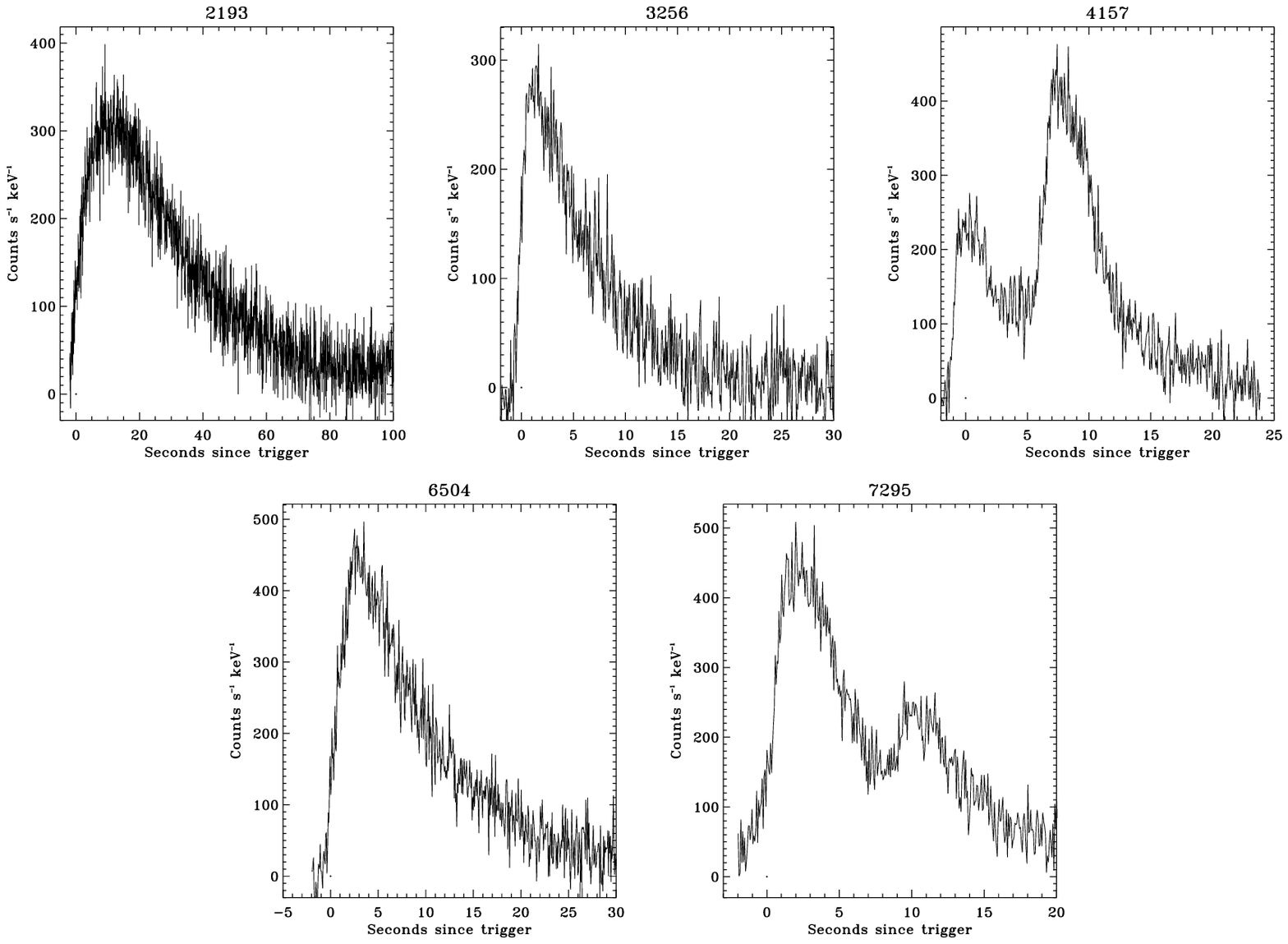}
 \caption{Light curves of the five thermal pulses used in the study.
 The pulses have hard spectra and are modelled
with black-body spectra.} \label{fig:lc1}
\end{figure}

\begin{figure}[]
\epsscale{0.8}
 \plotone{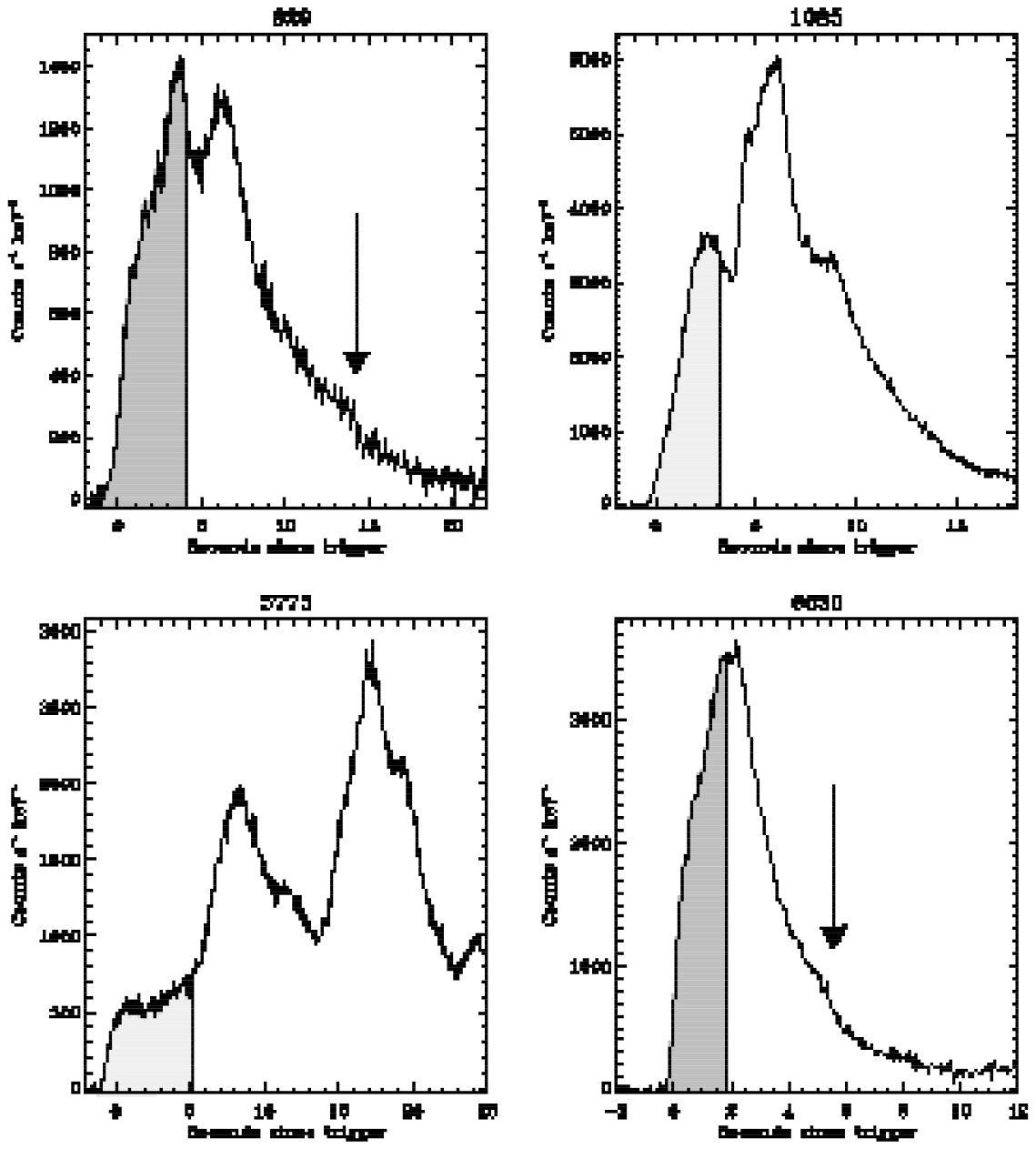}
 \caption{During the initial phase of these bursts, marked by the shaded
 areas, the spectra are consistent with a black body spectrum. The
 arrows draw  attention to the breaks in the light curve.
 See the text for details.} \label{fig:lc2}
\end{figure}

\begin{figure}[]
\epsscale{0.8}
 \plotone{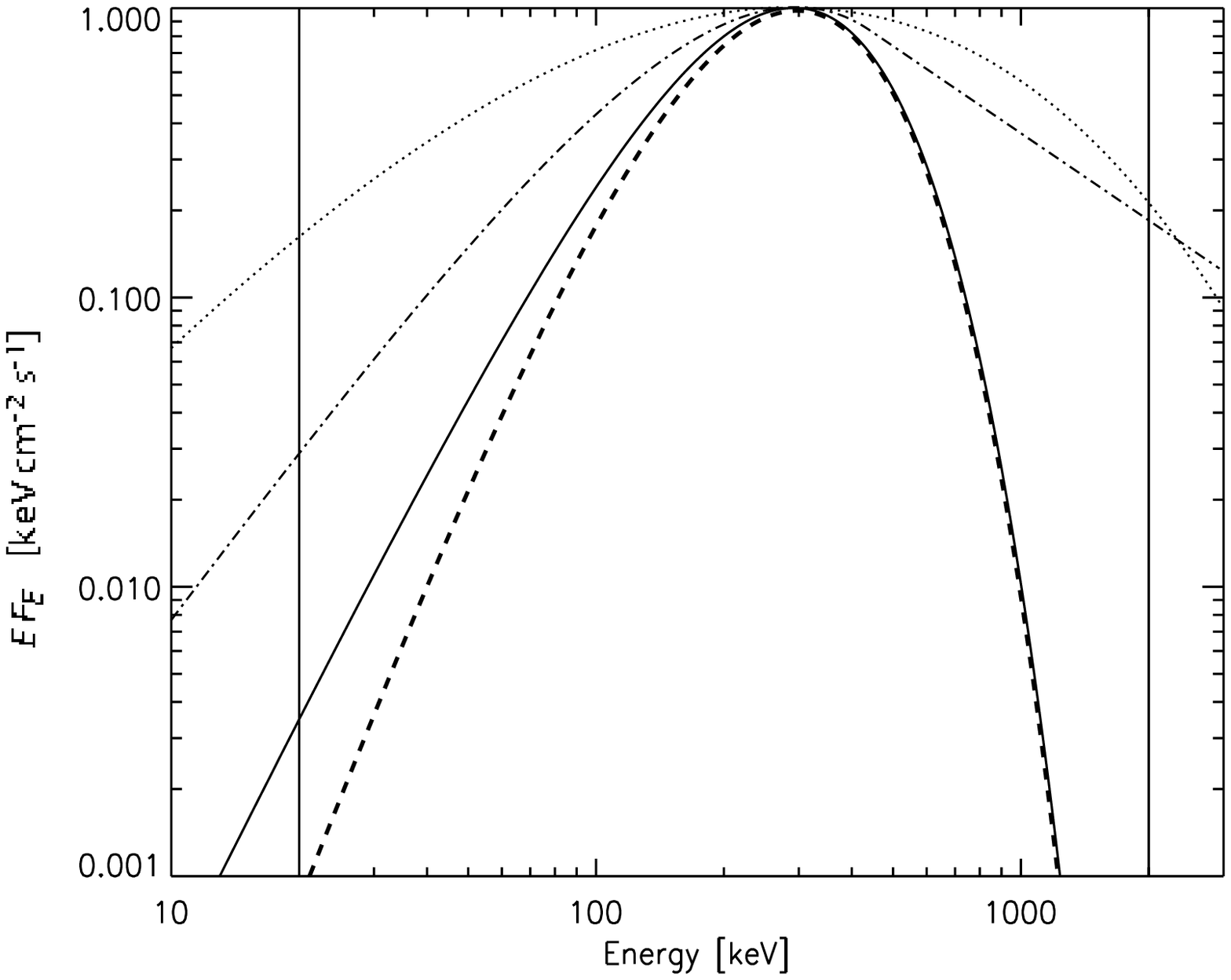}
\caption{Spectral distributions of various emissions with a hard,
low-energy, slope in a $E\, F_E$-representation; Planck spectrum
(solid line), Wien spectrum (dashed), optically-thin
thermal-synchrotron (dotted), and small-pitch angle synchrotron
emission from a soft energy-distribution of relativistic electrons
(dot-dashed). The latter spectrum can approximately represent a
Compton-attenuated synchrotron-spectrum, as well. The spectra are
normalized to each other to facilitate a comparison. The typical
BATSE energy window is also indicated.} \label{fig:planck}
\end{figure}

\begin{figure}[]
 \epsscale{1.0}
 \plotone{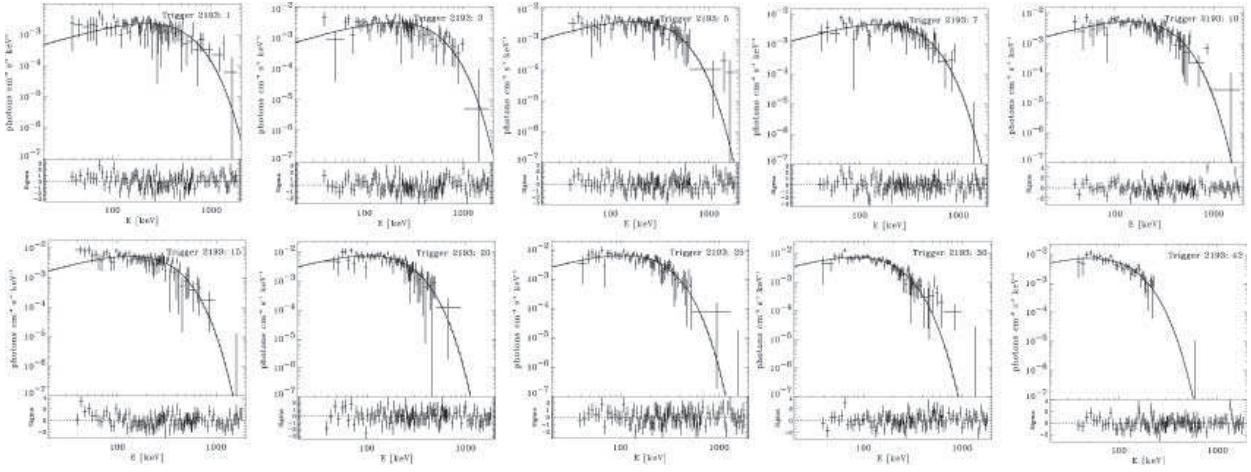}
 \caption{Black-body fits to the time-resolved spectra of burst
 2193.
The spectra correspond to the following time bins (compare fig.
[\ref{fig:lc1}]), 1:  1.6 -- 2.4 s, 3: 3.14 -- 3.84 s, 5:  4.5 --
5.3 s, 7: 6.0 -- 6.7 s, 10: 8.0 -- 8.7 s,
 15: 11.3 -- 11.9 s, 20: 14.6 -- 15.2 s , 25: 18.0 -- 18.8 s,
 30: 21.6 -- 22.9 s, 42: 39.4 -- 41 s.
 }
\label{fig:2193}
\end{figure}

\begin{figure}[]
 \epsscale{1.0}
 \plotone{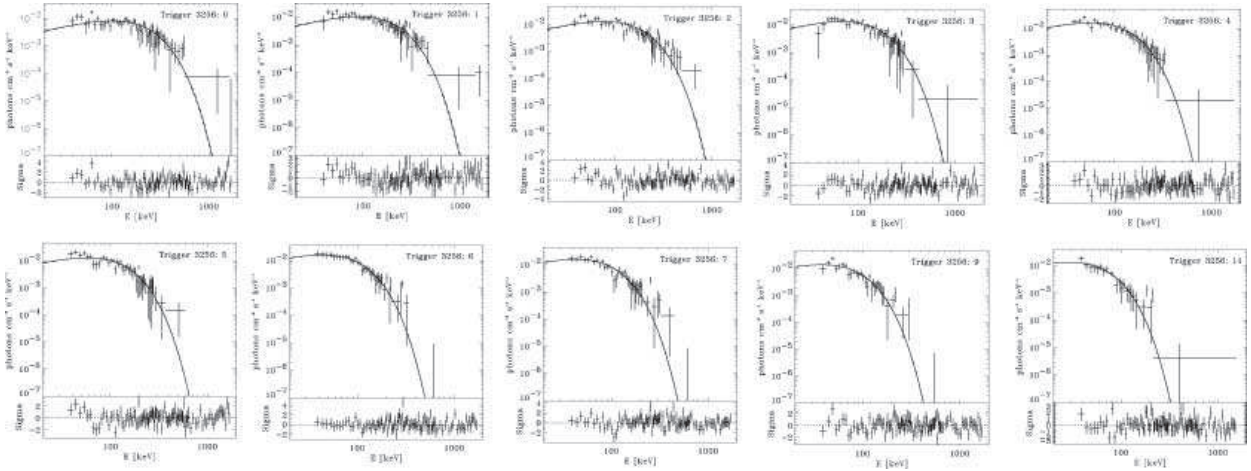}
 \caption{Same as fig. \ref{fig:2193} but for burst 3256, for time bins:
 0:0 -- 0.8 s , 1: 0.8 -- 1.5 s , 2: 1.5 -- 2.2 s, 3:2.2 -- 1.9 s,
 4:2.9 -- 3.6 s, 5: 3.6 -- 4.5 s, 6: 4.5 -- 5.4 s, 7: 5.4 -- 6.3 s, 9:
 7.4 -- 8.6 s, 11: 9.9 -- 11.3 s.
 }
\label{fig:3256}
\end{figure}

\begin{figure}[]
 \epsscale{1.0}
 \plotone{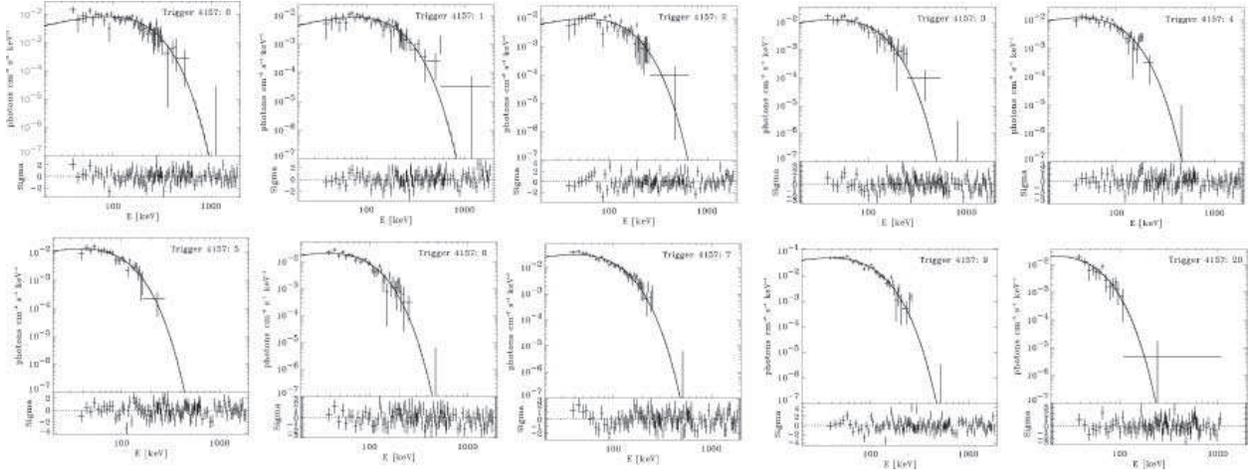}
 \caption{Same as fig. \ref{fig:2193} but for burst 4157, for time
 bins: 0: 0.1 -- 0.7 s , 1: 0.7 -- 1.5 s , 2: 1.5 -- 2.4 s, 3: 2.4 -- 3.3 s,
 4: 3.3 -- 4.4 s, 5: 4.4 -- 5.4 s, 6: 5.4 -- 6.3 s, 7: 6.3 -- 7.0 s ,
 9: 7.5 -- 8.1 s, 20: 18.6 -- 20.1 s.
 }
\label{fig:4157}
\end{figure}

\begin{figure}[]
 \epsscale{1.0}
 \plotone{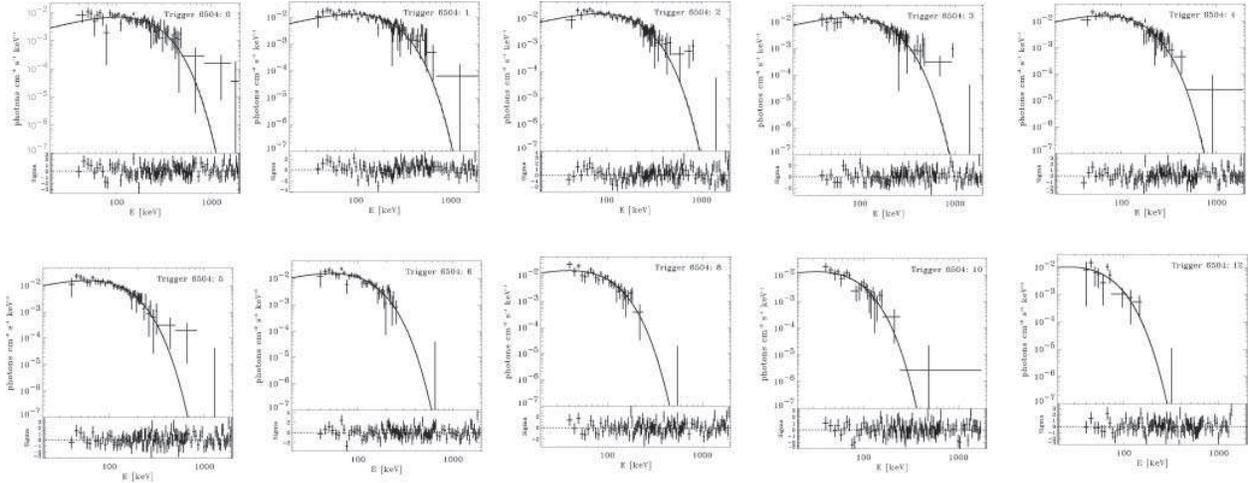}
 \caption{Same as fig. \ref{fig:2193} but for burst 6504, for time
 bins: 1: 1.6 -- 2.7 s, 2: 2.7 -- 3.4  s, 3: 3.4 -- 4.9  s, 4: 4.9 -- 6.2 s,
  5: 6.2 -- 7.7 s, 6: 7.7 -- 9.3 s, 8: 11.3 -- 13.4 s, 10: 15.9 -- 18.5 s,
  12: 21.4 -- 24.5 s.
 }
\label{fig:6504}
\end{figure}

\begin{figure}[]
 \epsscale{1.0}
 \plotone{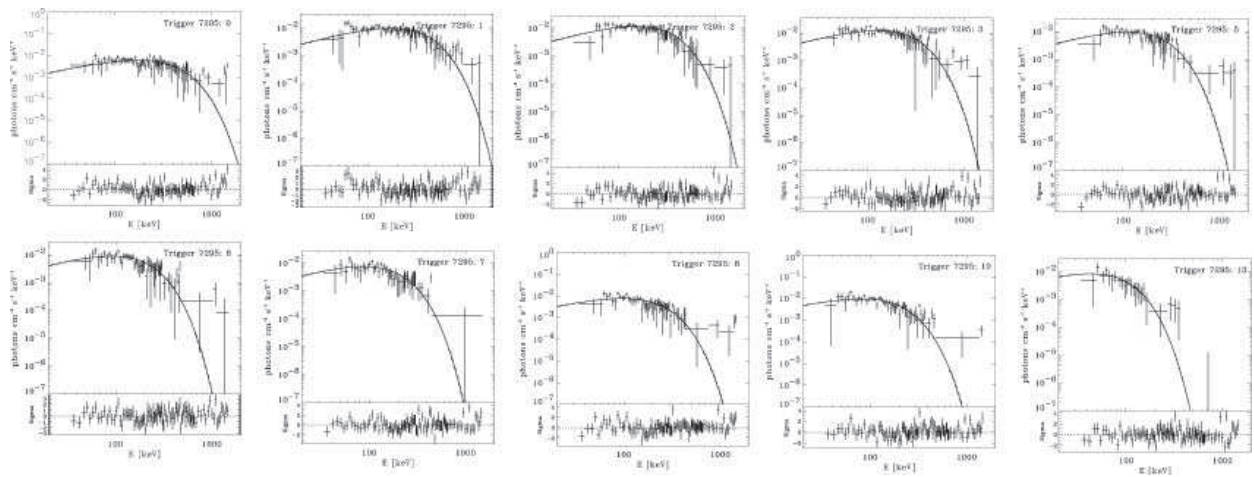}
 \caption{Same as fig. \ref{fig:2193} but for burst 7295, for time
 bins: 0: 0 -- 1.2 s, 1: 1.2 -- 2.1 s, 2: 2.1 -- 2.9 s, 3: 2.9 -- 3.8 s,
 5: 4.8 -- 5.9 s, 6: 5.9 -- 7.2 s, 7: 7.2 -- 8.6 s, 8: 8.6 -- 9.9 s,
 10: 11.1 -- 12.4 s, 13: 15.6 -- 17.7 s.
 }
\label{fig:7295}
\end{figure}

\begin{figure}[]
 \epsscale{0.9}
 \plotone{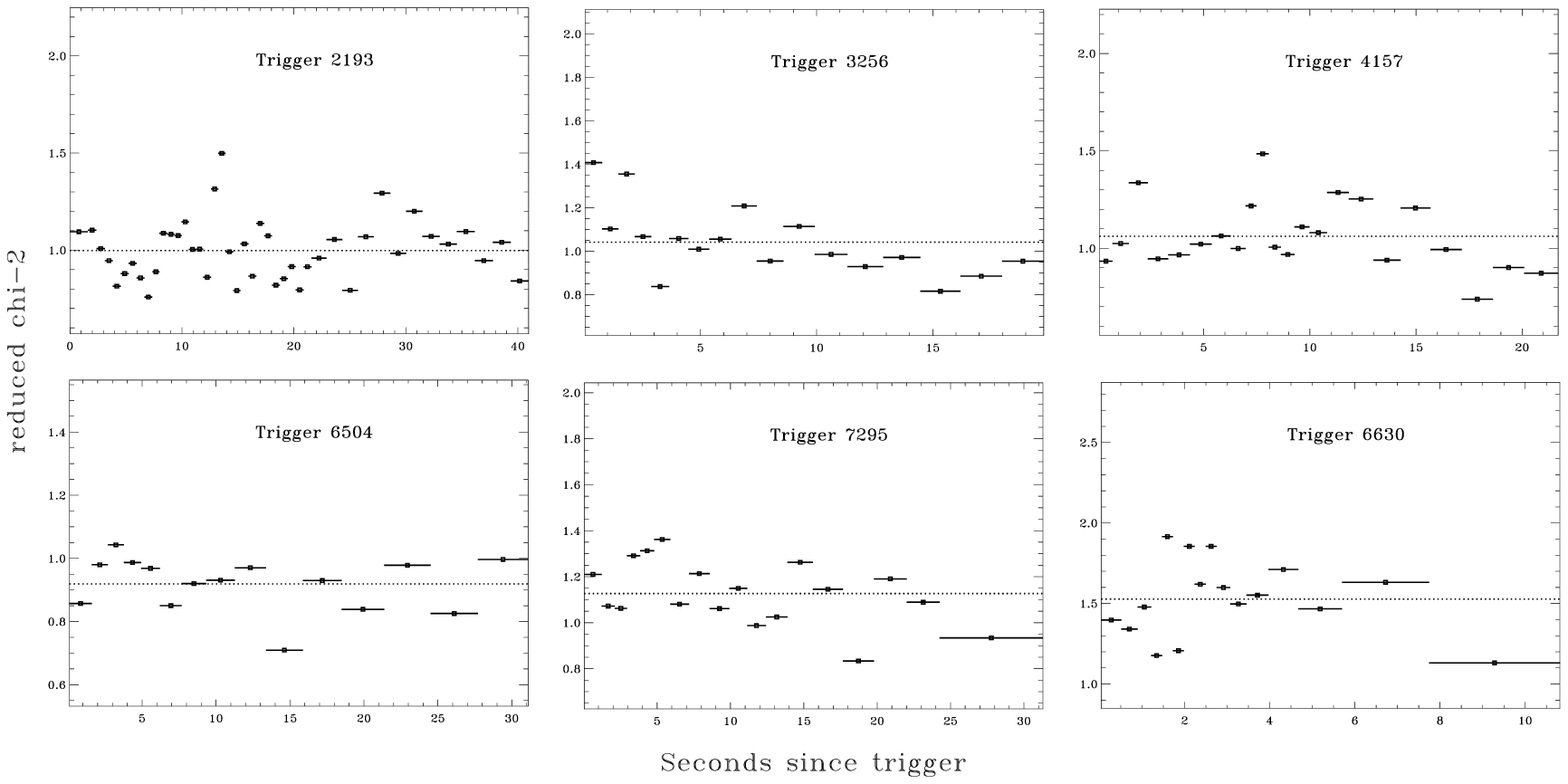}
 \caption{Reduced $\chi^2$-values, $\chi^2_\nu$, for the time-resolved spectra of the
 burst studied in the text. The dashed line is the reduced $\chi^2$-value for the total
 fit.} \label{fig:chi}
\end{figure}

\begin{figure}[]
 \epsscale{1.0}
 \plotone{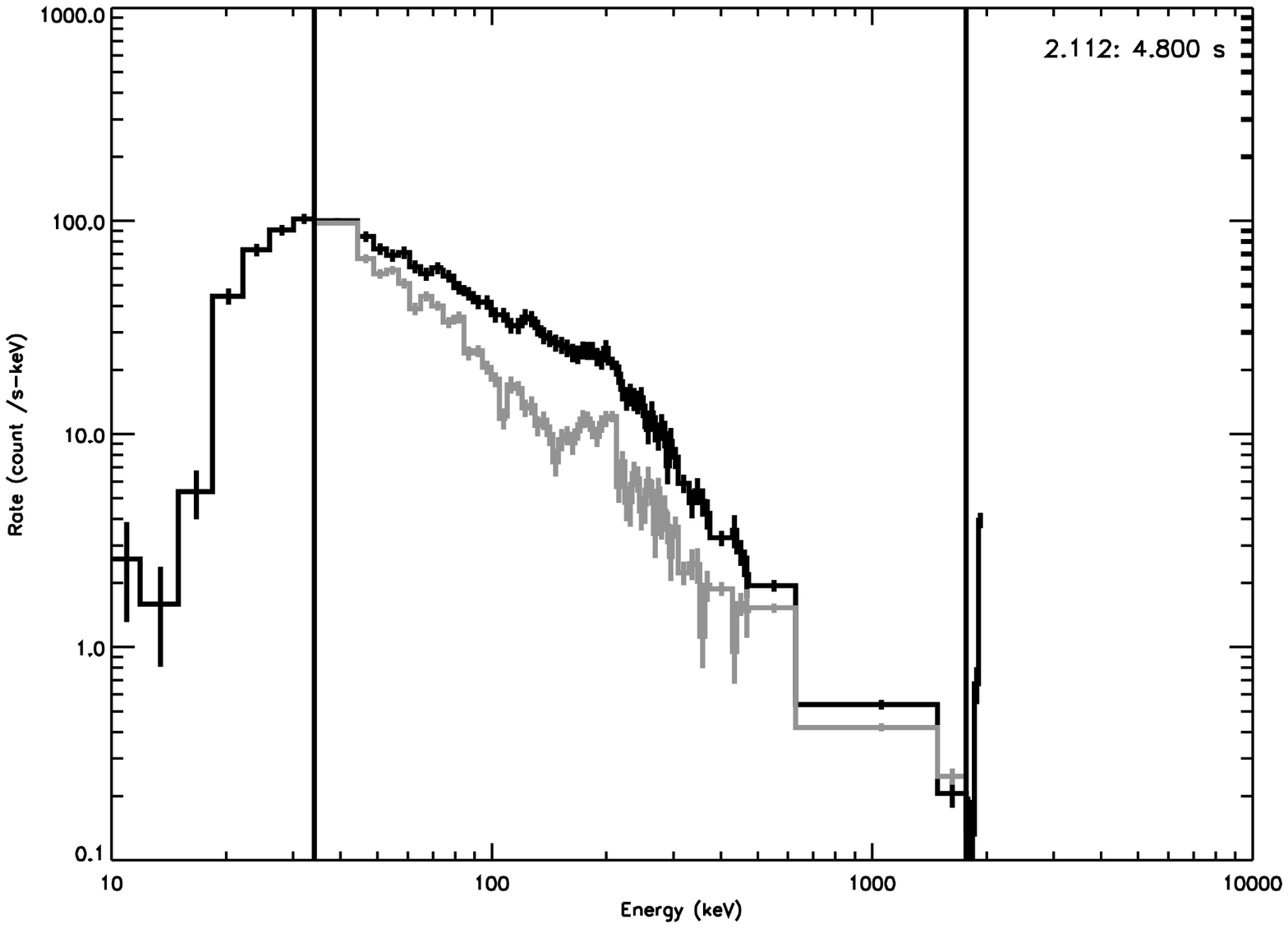}
 \caption{The BATSE raw data (solid black line), in counts s$^{-1}$ keV$^{-1}$ for
 the peak of bursts 7295, illustrating the background signal (grey line)
 compared to the GRB data. The studied interval is marked by the
 solid, vertical lines and the signal-to-noise ratio, SNR=1.
 }\label{fig:background}
\end{figure}

\begin{figure}[]
 \epsscale{0.9}
 \plotone{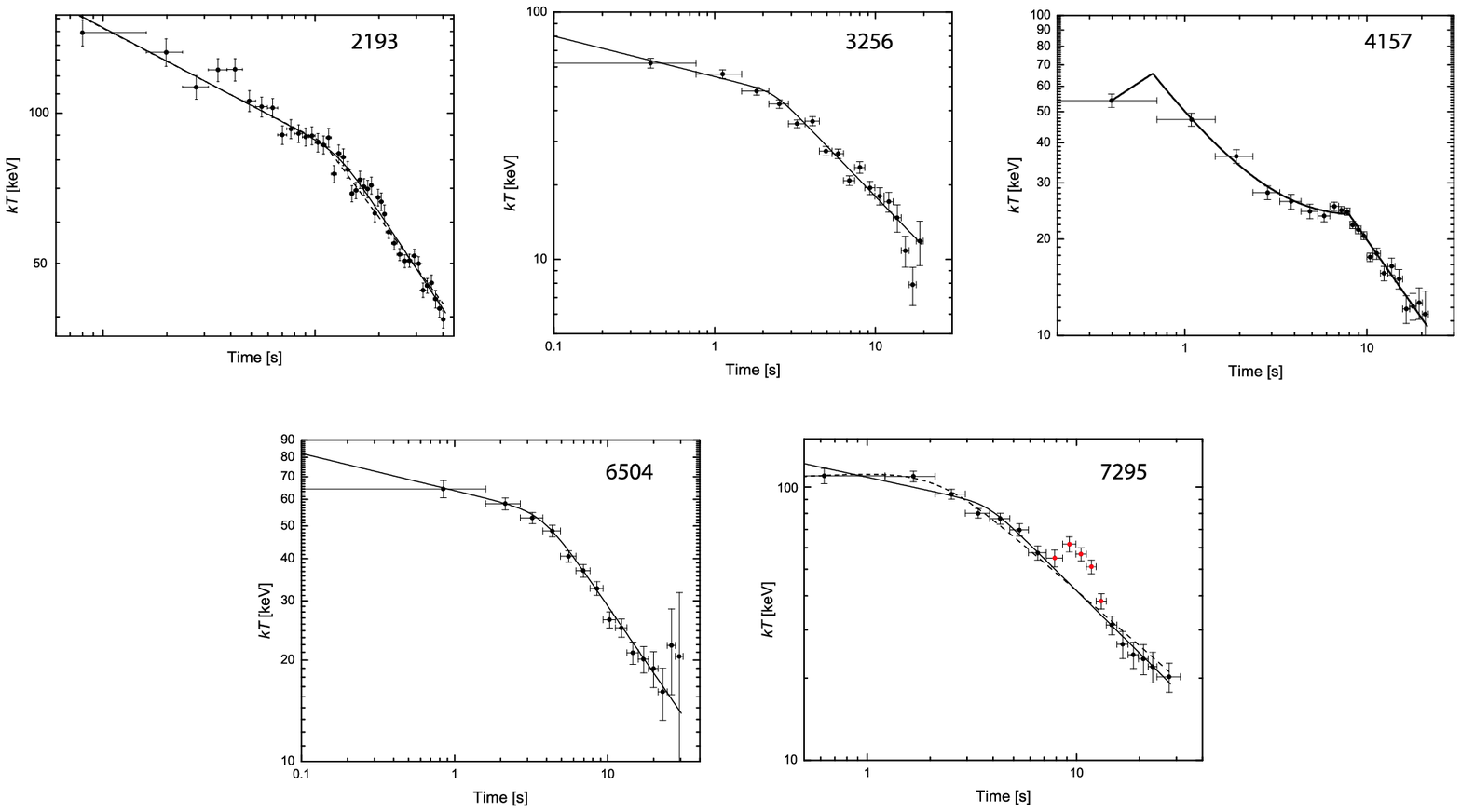}
 \caption{Black-body temperature, $kT$, in the observer frame, as a function
 of time. Single, broken, power-law functions (eq.[\ref{eq:smooth2}]) have been
 fitted to all pulses except trigger 4157, for which a sum of two, broken, power-laws
 has been fitted. The breaks in the figures correspond to the
 breaks in the light curves in Figure \ref{fig:lc1}.
 The solid lines correspond to fits with $\delta
 = 0.15$, while the dashed curves are the best fits with $b=
 -0.67$. See the text and Table \ref{tab:Tt} for details.
 \label{fig:BBkT} }
\end{figure}

\begin{figure}[]
 \epsscale{1.0}
 \plotone{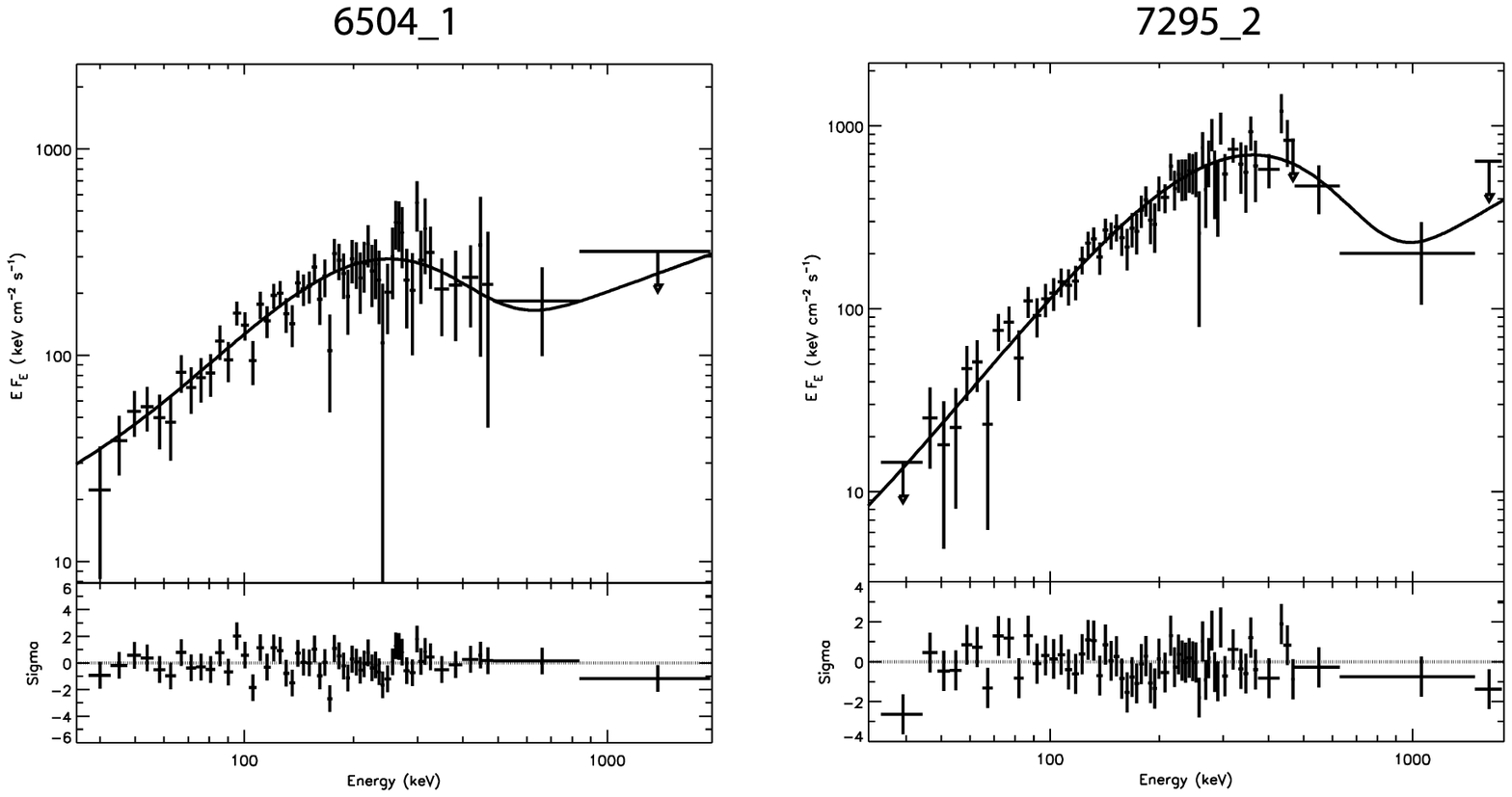}
 \caption{Combination of a black body and a single power-law fitted to time-resolved spectra
 of triggers 6504 (left panel) and 7295 (right panel). The digits
 following the trigger number correspond to the time bins in
 Figures \ref{fig:6504} and \ref{fig:7295}, respectively. Note that the plot is in
 an
 $EF_{E}$ representation.
 }\label{fig:extra}
\end{figure}

\begin{figure}[]
 \epsscale{1.0}
 \plotone{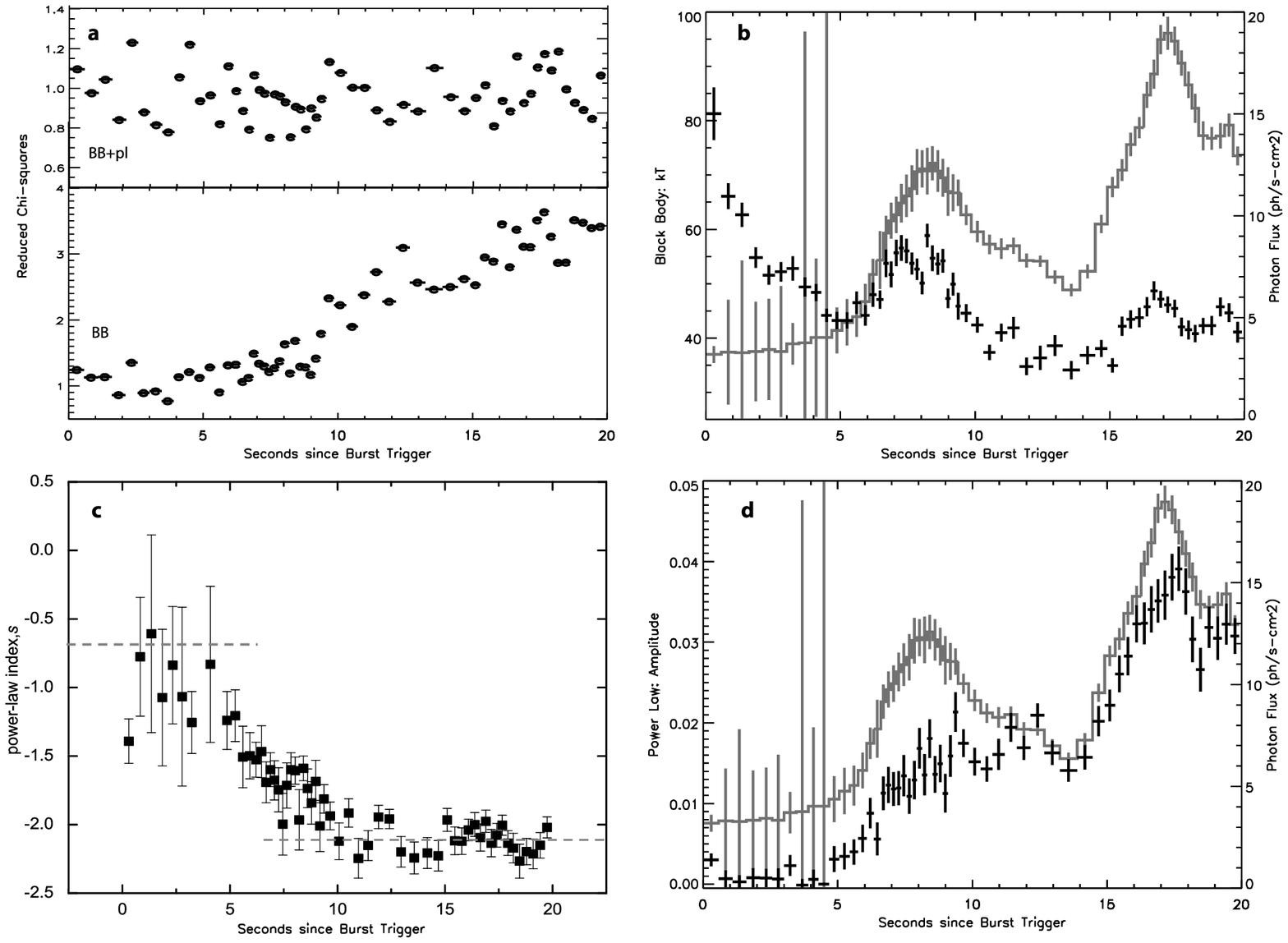}
 \caption{Analysis of trigger 5773. a) $\chi^2_\nu$ as a function
 of time for a black body (lower) and a black body and power law
 model (upper). b) Evolution of $kT$ plotted on top of the light curve
(grey line), with the time resolution used in the analysis. c) The
power-law index, $s$ as a function of time.  The dashed lines show
$s= -0.67$ (spectral-index for optically-thin synchrotron
emission) and $s= -2.1$ (cooling spectrum). d) The amplitude of
the power-law index grows in importance.
 }\label{fig:5773_4}
\end{figure}

\begin{figure}[]
 \epsscale{1.0}
 \plotone{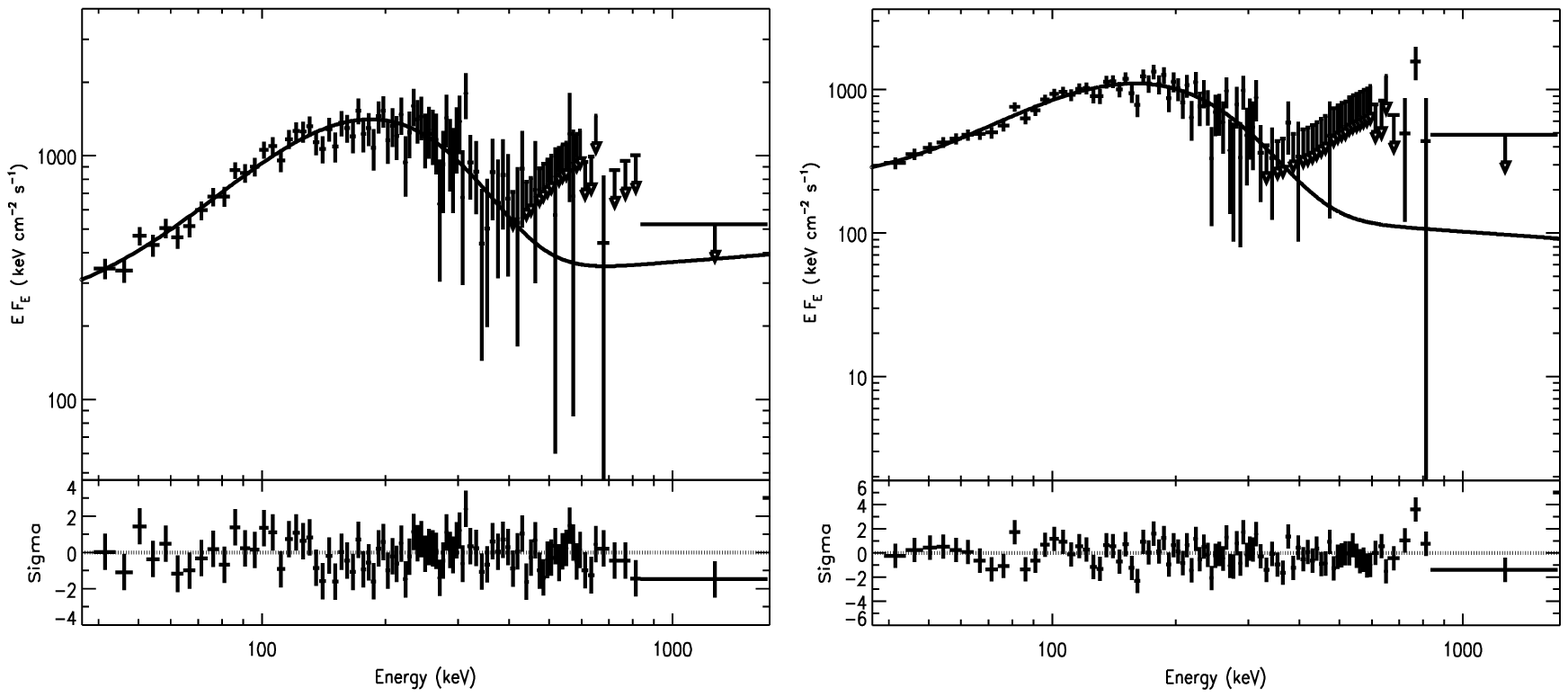}
 \caption{Time-resolved spectra of burst 6630 fitted with a two-component
 spectral model consisting of a black body and a power law. The time bins
 correspond to the times: $t = 2.50 -2.75$ s (left) and
 $t = 2.75 - 3.1$ s. The fits require a power law with index $s = -1.9 \pm 0.1$
 and $-2.2 \pm 0.2$, respectively. The upper-limit data-point at high
 energies represents the detector sensitivity.
 }\label{fig:6630}
\end{figure}

\end{document}